\documentclass[reprint,amsmath,amssymb]{revtex4-1}
\usepackage{etoolbox}

\usepackage{pdfpages}

\usepackage{etoolbox} 

\makeatletter
\patchcmd{\@outputpage@head}{\@ifx{\LS@rot\@undefined}{}{\LS@rot}}{}{}{}
\makeatother

\usepackage{upgreek}
\usepackage{graphicx}
\usepackage{soul}
\usepackage{braket}
\usepackage{lineno}
\begin{document}

\title{High-frequency cavity optomechanics using bulk acoustic phonons}

\author{P. Kharel$^{1}$}
\email{prashanta.kharel@yale.edu}

\author{G. I. Harris$^{2}$}

\author{E. A. Kittlaus$^{1}$}

\author{W. H. Renninger$^{1}$}

\author{N. T. Otterstrom$^{1}$}

\author{J. G. E. Harris$^{2}$}

\author{P. T. Rakich$^{1}$}

\email{peter.rakich@yale.edu}

\affiliation{$^{1}$ Department of Applied Physics, Yale University, New Haven, Connecticut 06511, USA}

\affiliation{$^{2}$ Department of Physics, Yale University, New Haven, Connecticut 06520, USA}

\begin{abstract}
To date, micro- and nano-scale optomechanical systems have enabled many proof-of-principle quantum operations through access to high-frequency (GHz) phonon modes that are readily cooled to their thermal ground state. However, minuscule amounts of absorbed light produce excessive heating that can jeopardize robust ground state operation within such microstructures. In contrast, we demonstrate an alternative strategy for accessing high-frequency ($13$ GHz) phonons within macroscopic systems (cm-scale). Counterintuitively, we show that these macroscopic systems, with motional masses that are $>20$ million times larger than those of micro-scale counterparts, offer a complementary path towards robust quantum operations. Utilizing bulk acoustic phonons to mediate resonant coupling between two distinct modes of an optical cavity, we demonstrate the ability to perform beam-splitter and entanglement operations at MHz rates on an array of phonon modes, opening doors to applications ranging from quantum memories and microwave-to-optical conversion to high-power laser oscillators. 
\end{abstract}

\maketitle
\section*{Introduction}
The coherent control of mechanical objects \cite{cirac1995quantum, o2010quantum, teufel2011circuit, chu2017quantum} can enable applications ranging from sensitive metrology \cite{stowe1997attonewton, krause2012high} to quantum information processing \cite{leibfried2003quantum,stannigel2010optomechanical, lee2012macroscopic}.
An array of devices \cite{metzger2004cavity, schliesser2006radiation, arcizet2006radiation, regal2008measuring,  brennecke2008cavity, thompson2008strong,eichenfield2009optomechanical,ding2010high,bahl2012observation} ranging from nano-optomechanical crystals to suspended micro-mirrors have been used to manipulate mechanical degrees of freedom using light. 
Central to the field of cavity optomechanics---and more generally to quantum information science---is the ability to harness long-lived mechanical excitations at high frequencies \cite{aspelmeyer2014cavity}.
Long-lived high-frequency phonons can be initialized deep in their quantum ground states and can preserve coherent information for extended periods of time in the presence of various decoherence channels. 
%
%
Optomechanical systems utilizing high-frequency (GHz) phonons have enabled ground state cooling \cite{chan2011laser}, quantum control at the single phonon level \cite{cohen2015phonon,riedinger2016non, hong2017hanbury}, and remote entanglement between mechanical resonators \cite{lee2011entangling,riedinger2018remote, ockeloen2018stabilized}.
%
%
%
%
However, it may be difficult to implement more sophisticated quantum protocols using these systems because of spurious forms of laser heating that continue to threaten robust ground-state operation within micro- and nano-scale systems \cite{arcizet2009cryogenic, chan2011laser, hong2017hanbury}.
%
%
New device strategies may be necessary to mitigate such decoherence mechanisms for robust quantum optical control of high-frequency phonons.


\begin{figure*}[]
\centerline{
\includegraphics[width=\linewidth]{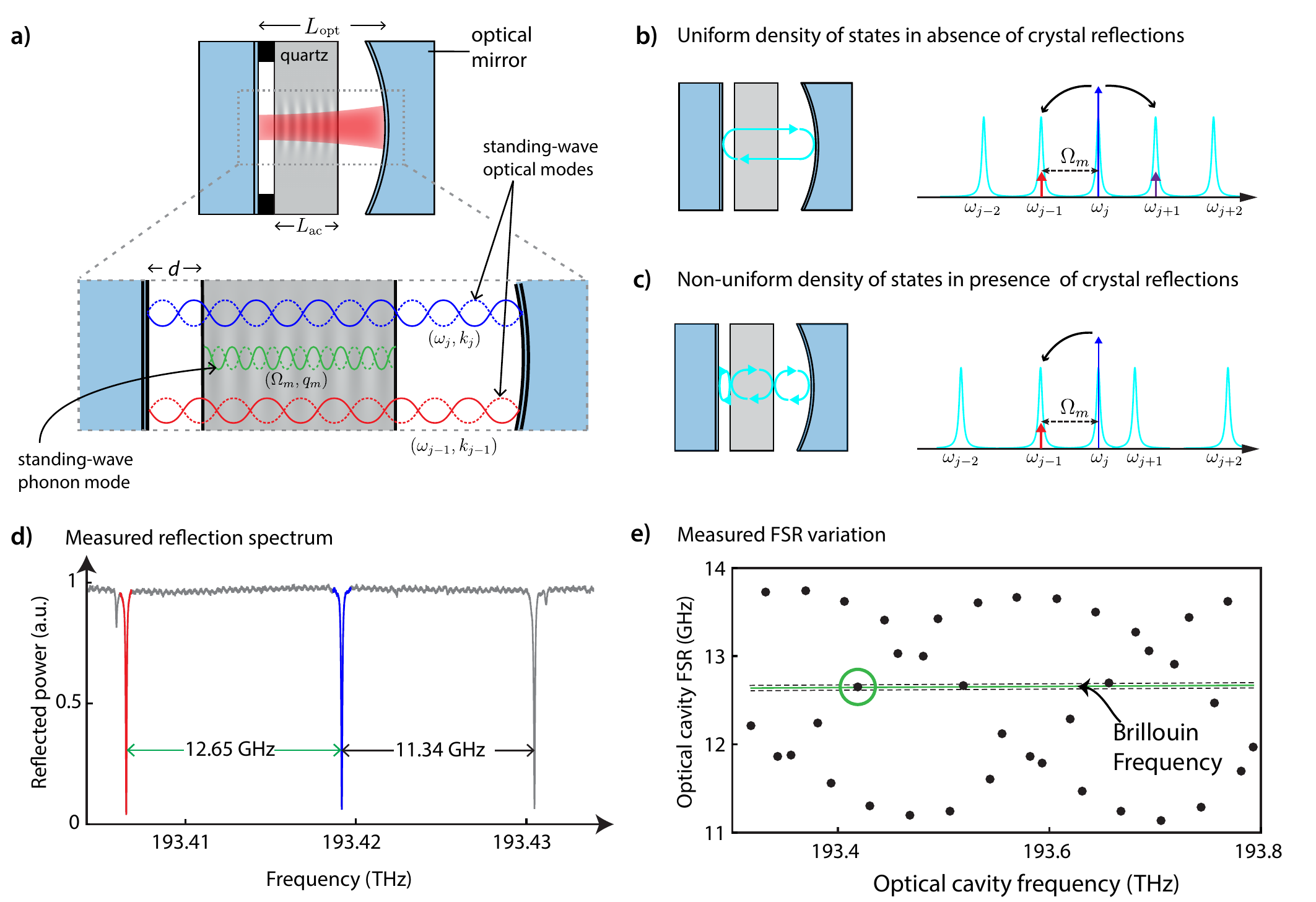}}
\caption{ \textbf{The multi-mode cavity optomechanical system.} \textbf{a}, Schematic of an optomechanical system that consists of a bulk acoustic wave resonator that is placed inside an optical cavity.
Two distinct standing-wave longitudinal optical modes  interact through electrostrictive coupling with a standing-wave longitudinal phonon mode formed within the planar crystalline quartz crystal at cryogenic temperatures ($\sim$ 8 K).
Large mode volumes for both light and sound result from macroscopic geometrical parameters: $L_\text{ac} \simeq 5.2$ mm, $L_\text{opt} \simeq$ 9.1 mm, and $d \simeq 0.15$ mm. 
%
%
\textbf{b}, Assuming no optical reflections in the quartz-vacuum interface, one obtains equally spaced longitudinal optical modes. Therefore, a phonon mode mediating inter-modal optomechanical interactions can scatter incident laser light at $\omega_j$ to adjacent optical modes $\omega_{j+1}$ and $\omega_{j-1}$ with nearly equal scattering rates.  
%
\textbf{c}, However, modest optical reflections ($\sim 4\%$) in the quartz-vacuum interface, which lead to significant non-uniformity in optical mode spacing, can be exploited to bias our system to strongly favor one scattering process ($\omega_j \rightarrow \omega_{j-1}$) over the other ($\omega_j \rightarrow \omega_{j+1}$). The variation in cavity resonances arises because of dispersive shifts of cavity resonances arising from multi-path interference. 
%
%
\textbf{d}, The reflection spectrum of the optical cavity obtained by frequency-sweeping an incident laser light reveals this variation in optical mode spacing (or free spectral range) between adjacent fundamental Gaussian modes. 
\textbf{e}, Measurement of the free spectral range (FSR) over a wider frequency range by sweeping the laser from wavelength 1548 to 1552 nm shows large undulation (2.6 GHz) in the FSR as a function of the optical cavity frequency.
Consequently, we find multiple pairs of resonances with frequency differences that are equal to the Brillouin frequency, which is a necessary requirement for inter-modal optomechanical coupling in our system.
}
\label{fig1}
\end{figure*}  
\color{black}
%
%
Efficient optomechanical coupling to high-frequency bulk acoustic waves within macroscopic systems offer an alternative path to robust quantum control of phonons. %
%
Optical access to such phonons within pristine crystalline solids yield greatly reduced surface interactions and favorable thermal characteristics to mitigate spurious laser heating \cite{galliou2013extremely, renninger2018bulk}.
%
%
These resonators also grant access to world record  $f\!\cdot\! Q$-products \cite{galliou2013extremely},  a key figure of merit that characterizes decoupling of a resonator form its thermal environment  \cite{aspelmeyer2014cavity}.
%
%
%
While such bulk acoustic phonon modes have been accessed through { Brillouin-like coupling using a free-space laser beam \cite{renninger2018bulk, kharel2018ultra}}, new strategies are needed to translate this physics into cavity optomechanical systems as the basis for phonon counting \cite{vanner2013quantum, cohen2015phonon}, generation of non-classical mechanical states \cite{hong2017hanbury}, and efficient transduction of information between optical and phononic domains \cite{parkins1999quantum, palomaki2013entangling, reed2017faithful}.

\raggedbottom
In this paper, we realize a cavity optomechanical system that harnesses high-frequency (13 GHz) bulk acoustic phonon modes within a macroscopic crystal, offering an array of properties that open the door to robust quantum manipulation of phonons. 
These high-frequency phonons, with large ($20\ \upmu$g) motional masses, are used to mediate resonant coupling between two distinct modes of an optical cavity through Brillouin-like interactions.
%
To enable beam-splitter and entanglement interactions as the basis for quantum optical control of phonons, we engineer the multi-mode spectrum of the optical cavity to break the symmetry between the Stokes and the anti-Stokes processes while allowing resonant driving of a chosen optical mode.
%
%
Resonant driving allows us to attain intra-cavity photon numbers that are more than $6$-orders of magnitude larger than those of high-frequency micro-scale counterparts \cite{chan2011laser}. 
These resonantly enhanced photon numbers permit large (MHz) optomechanical coupling rates and greater than unity co-operativities, necessary for efficient control of phonons using light.
%
%
In addition, we demonstrate that phase-matched Brillouin interaction produces controllable coupling to one or more phonon modes, opening the door to new forms of multi-mode entanglement \cite{vitali2007stationary, woolley2014two, weaver2018phonon}.
%
%
Looking beyond the field of cavity quantum optomechanics, this device strategy presents new opportunities for sensitive materials spectroscopy and oscillator technologies.

\section*{Bulk Crystalline Optomechanical System}

%
In what follows, we explore optomechanical interactions mediated by macroscopic phonon modes within the bulk crystalline cavity optomechanical system of  Fig. \ref{fig1}a. 
A flat-flat quartz crystal, of 5.2 millimeter thickness, is placed within a nearly hemispherical Fabry-P\'erot cavity having high  reflectivity (98\%) mirrors; this optomechanical assembly is cooled to $\sim 8$ Kelvin temperatures to greatly extend the phonon lifetimes within crystalline quartz.
%
%
%
At room temperature, high-frequency phonons ($>$10 GHz) have a mean-free path ($\sim 100 \ \upmu$m) that is much smaller than the crystal dimension.
Cooling this system to cryogenic temperatures, extends the phonon coherence to meter length-scales, permitting the formation of macroscopic high-frequency phonon modes \cite{ohno2015spectral, renninger2018bulk}.
%
%
Just as the optical Fabry-P\'erot resonator supports a series of standing electromagnetic waves (red, blue) seen in Fig. \ref{fig1}a, the planar surfaces of the quartz crystal produce acoustic reflections to form an acoustic Fabry-P\'erot that supports a series of standing-wave elastic modes (green) at these low temperatures.
%
%
To avoid anchoring losses, an annular spacer prevents the crystal surface from contacting the mirror; this same spacer sets the position of the crystal ($d$) within the cavity (See Methods and Supplementary Note: I for further details). 


%
These high-frequency bulk acoustic phonon modes are used to mediate efficient coupling between distinct longitudinal optical modes of the Fabry-P\'erot through a multi-resonant (or multi-mode) optomechanical process.
%
%
Coupling occurs through a Brillouin-like optomechanical process when the time-modulated electrostrictive optical force distribution, produced by the interference between distinct modes of the Fabry-P\'erot, matches the elastic profile (and frequency) of a bulk acoustic phonon mode.
The same intrinsic photoelastic response that generates the optical forces within the crystal also modulates the refractive index of the crystal via elastic-wave motion.
Through the formation of a time-modulated photoelastic grating, such `Brillouin-active' elastic waves mediate scattering (or energy transfer) between distinct longitudinal modes of the optical Fabry-P\'erot cavity.

%
%
Because of the extended nature of the optomechanical interaction within the crystal, phase-matching and energy conservation determine the set of phonon modes ($\Omega_m, q_m$) that can mediate resonant coupling between adjacent optical modes ($\omega_j,k_j$)  and ($\omega_{j-1},k_{j-1}$) of the Fabry-P\'erot (See Fig. \ref{fig1}a).
One finds that the Brillouin active phonons must satisfy the conditions $q_m = k_{j-1} +k_j$ and $ \Omega_m = \omega_j-\omega_{j-1}$ in order to mediate dynamical Bragg scattering within the crystal.
%
%
Neglecting some details pertaining to modal overlaps, one expects that phonons in a narrow band of frequencies near the Brillouin frequency ($\Omega_B \simeq 2\omega_j v_a/v_o$) can meet this condition;
here, $v_a$ ($v_o$) is the speed of sound (light) in the quartz crystal.
Note that this resonance condition is dictated entirely by the frequency of light ($\omega_j$) and intrinsic properties of the crystal.
%
%
Within the $z$-cut crystalline quartz substrate, this expression leads us to expect inter-modal coupling at $\Omega_B \simeq 2\pi \times 12.7$ GHz when driving the optical cavity with $1.55 \ \upmu$m wavelength light.
Based on the resonance condition, we seek a Fabry-P\'erot design whose free spectral range (FSR), defined as $\omega_\text{FSR}^j = \omega_{j}-\omega_{j-1}$, matches the Brillouin frequency $\Omega_B$.  

Most applications of optomechanical interactions require a means of selecting between Stokes or anti-Stokes interactions.
%
%
Conventional single-mode cavity optomechanical systems use the detuning of an external drive field from a single cavity resonance to produce this asymmetry.
%
%
By comparison, this multi-mode system offers the possibility for resonant pumping, which carries many advantages (See Discussion).
However, in the case when the optical Fabry-P\'erot resonator has regular mode spacing (i.e., $\omega_\text{FSR}^j = \omega_\text{FSR}^{j-1} $), resonant driving of the optical cavity presents a problem; a Brillouin-active phonon mode ($\Omega_m$) that matches the {multi-mode resonant condition} will resonantly scatter incident photons of frequency $\omega_{j}$ to adjacent cavity modes $\omega_{j-1}$ and $\omega_{j+1}$ with nearly equal probabilities (See Fig. \ref{fig1}b). 
The introduction of the quartz crystal into the optical Fabry-P\'erot provides an elegant means of solving this problem; modest optical reflections ($4\%$) produced by the surfaces of the crystal shift the modes of the Fabry-P\'erot. As a result, the typically uniform density of modes (Fig. \ref{fig1}b) is transformed into a highly non-uniform density of modes (Fig. \ref{fig1}c) such that $\omega_\text{FSR}^j \neq \omega_\text{FSR}^{j-1} $. Using this strategy, we are able to choose between Stokes ($\omega_{j}\rightarrow \omega_{j-1} $) or anti-Stokes ($\omega_{j-1}\rightarrow \omega_{j} $) processes with high selectivity, even when the external drive field is directly on resonance. For example, Fig. \ref{fig1}c illustrates how a dispersively engineered mode spacing ($\omega_\text{FSR}^j \neq \omega_\text{FSR}^{j-1} $) permits us to bias the system for Stokes scattering with resonant optical driving at frequency $\omega_{j}$.

%
This modification to the density of modes can be seen from reflection measurements of this bulk crystalline optomechanical system (Fig. \ref{fig1}d), which reveal a large ($\sim 21\%$) undulation in the FSR of the optical Fabry-P\'erot; figure \ref{fig1}e shows the measured frequency spacing  between adjacent optical resonances ($\omega_\text{FSR}^j = \omega_{j+1}-\omega_j$) when an incident laser field is mode-matched to the fundamental Gaussian mode of the cavity. 
The observed variation in optical FSR agrees well with scattering matrix treatments of the system (See Supplementary Notes: I).  
%
%
The large spread in mode separations makes it straightforward to find pairs of optical modes whose frequency difference satisfies the multi-mode resonance condition.
Within the measurements of Fig. \ref{fig1}e, one can readily identify three sets of modes (green) that satisfy this condition. 
Moreover, this dispersive coupling permits us to fine-tune the FSR of a given mode pair to match Brillouin frequency through small ($<$ 1 Kelvin) change in temperature (See Supplementary Notes: I).

Even with resonant driving of the optical mode, this form of symmetry breaking results in large ($>1000$-fold) difference between the Stokes and the anti-Stokes scattering rates.
%
%
In conventional single-mode cavity optomechanical systems, such large Stokes/anti-Stokes asymetry is produced by far-detuning the drive from an optical cavity at the expense of intracavity photon number.
In contrast, this multi-mode optomechanical system permits greatly enhanced intracavity photon numbers for the same input power through resonant driving.
The relative strength of the Stokes/anti-Stokes scattering rates in the case of resonant driving is determined by the ratio $(2 \Delta \omega/ \kappa)^2$, where $\Delta \omega = \omega_\text{FSR}^{j+1} -\omega_\text{FSR}^j$ is the difference in the FSR between then adjacent optical modes and $\kappa$ is the optical mode linewidth (See Supplementary Notes: VII for details).
Because $\Delta \omega$ is well-resolved from the linewidth (i.e., $2\Delta \omega/\kappa \simeq 36$) we can virtually eliminate the Stokes or anti-Stokes interaction by resonantly exciting an appropriately chosen mode.

\begin{figure}[]
\centerline{
\includegraphics[width=\linewidth]{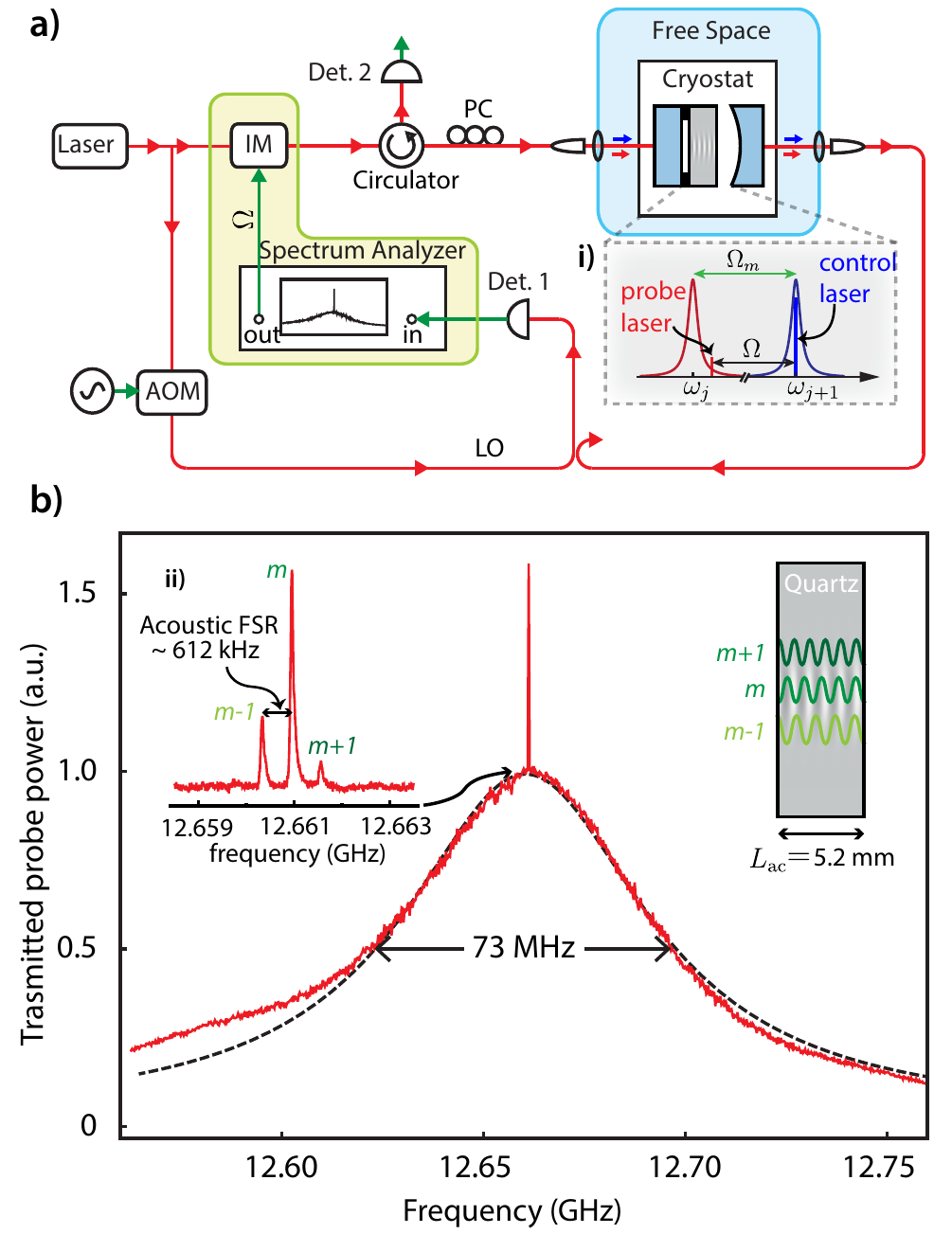}}
\caption{\textbf{Optomechanical coupling to a multitude of phonon modes.} 
\textbf{a}, Schematic of the measurement setup used to perform optomechanically induced amplification (OMIA).
A strong control and a weak probe laser is synthesized from the same tunable laser source using an intensity modulator (IM). 
This light is coupled into and out of the optical cavity using fiber-optic collimators and free-space lenses.
The frequency of the tunable laser is set so that the control laser is directly on resonance with the high-frequency optical mode at  $\omega_{j+1}$ (See inset \textbf{i}).
The probe laser is swept near the lower cavity mode at frequency $\omega_j$ by sweeping the RF-drive frequency ($\Omega$) of the intensity modulator.
To examine the coherent response of the intra-cavity probe field due to optomechanical coupling, a heterodyne measurement is performed between the transmitted probe light and acousto-optic modualtor (AOM) shifted laser light. 
\textbf{b}, We observe coherent build up of intra-cavity photon number as the probe laser is scanned near the optical mode $\omega_j.$ The optical cavity decay rate of $\kappa_j \simeq 2\pi \ \times 73 $ MHz obtained for this mode is consistent with the losses at the two mirrors. 
Additionally, we see sharp resonances on the optical mode spectrum corresponding to the phonon mediated transfer of energy from the high-frequency optical mode at frequency $\omega_{j+1}$ to the lower-frequency optical mode at $\omega_{j}$.
A close zoom in of the optical mode spectrum near the center of the optical resonance reveals three high-frequency acoustic modes around $12.661$ GHz with frequency spacing of $\sim$ 612 kHz (See inset \textbf{ii}).
This frequency spacing is consistent with the acoustic free spectral range of $v_a/2L_\text{ac}$ for the standing-wave longitudinal modes formed in crystalline quartz along $z$-axis.
These acoustic modes have very high longitudinal mode numbers (or overtone number) of $m \approx 2.08\times 10^4$. 
}
\label{fig3}
\end{figure}

%
%
The multi-mode coupling described above can be represented by the interaction Hamiltonian $\hat{H}_\text{int}^m = -\hbar g_0^m (\hat{a}_{j+1}^{\dagger}\hat{a}_j \hat{b}_m+\hat{a}_j^{\dagger}\hat{a}_{j+1}\hat{b}_m^{\dagger})$ \cite{borkje2012quantum}.
%
Here, $\hat{a}_j^{\dagger}$ is the creation operator for the optical mode at frequency $\omega_j$, $\hat{b}_m^{\dagger}$ is the creation operator for the phonon mode at frequency $\Omega_m$, and $g_0^m$ is the single-photon coupling rate that characterizes the strength of the optomechanical interaction (see Supplementary Notes: II.A-C). 
This single-photon coupling rate characterizes the change in the frequency spacing between the two optical modes resulting from the dynamical modulation of the refractive index of the crystal.
Since the acoustic FSR ($v_a/2L_\text{ac}$) of the bulk acoustic waves within our quartz substrate is much smaller than the optical linewidth ($\kappa/2\pi$), more than one phonon mode ($\Omega_m$) near the Brillouin frequency ($\Omega_B$) can participate in the optomechanical coupling.
Hence, the total interaction Hamiltonian, which includes contributions from all these phonon modes, becomes $\hat{H}_\text{int}= \sum_m \hat{H}_\text{int}^m.$ 

The single-photon coupling rate, $g_0^m$, is calculated from the overlap integral of the normalized mode profiles for the optical and the acoustic fields:
\begin{align}
    g_0^m &\approx \frac{2 n^3 p_{13} \omega_j^2}{cL_\text{opt}}\sqrt{\frac{\hbar}{AL_\text{ac}\rho \Omega_m}} \times \nonumber \\  \label{g0m}
    &\int_d^{d+L_\text{ac}} \text{d}z \ \text{sin}(k_{j+1}z) \ \text{sin}(k_jz) \ \text{sin}(q_m (z-d)).
\end{align}
Here $n$, $\rho$, and $p_{13}$ are the refractive index, mass density, and photoelastic coefficients of alpha quartz, $c$ is the speed of light in vacuum, $L_\text{opt} $ is the length of the optical cavity, $L_\text{ac}$ the length of the crystal, $A$ is the effective area of the optical mode, and $k_j$ ($q_m$) is the optical (acoustic) spatial period of standing-wave optical (acoustic) modes within quartz (See Supplementary Notes: II.D).
Using the parameters in Methods, Eq. (\ref{g0m}) predicts a single-photon coupling rate of $g_0^m \simeq 2\pi \times 20$ Hz. Note that this coupling rate is produced by a phonon mode with a motional mass of 20 $\upmu$g, which is nearly a million times larger than prior GHz-frequency optomechanical systems \cite{chan2011laser,ding2010high, mitchell2016single} (See Methods for details). 

Interestingly, the wavevector-selective nature of this coupling enables new approaches to precisely tailor interaction with one or more phonon modes.
This system differs from conventional Brillouin scattering because even phonon modes that satisfy both the energy conservation and the phase-matching requirements can have vanishing optomechanical coupling rates.
This intriguing new feature arises because the coupling to a particular phonon mode also depends on the location of the crystal ($d$) inside the optical cavity (See equation (\ref{g0m})).
For instance, the overlap integral in equation (\ref{g0m}) yields zero-coupling rate when the position of the crystal is such that nodes of a phonon mode coincide with the anti-nodes of the optical forcing function.
%

Despite the fact that this current apparatus does not permit us to change the crystal position, we can still control the number of phonon modes that participate in the optomechanical interaction. 
%
%
Independent of crystal location, a pair of adjacent optical modes are guaranteed to couple to at least one phonon mode near the Brillouin frequency because the phase-matching condition is relaxed by the crystal's finite length.
However, by using optical resonances at different wavelengths, one can change the position of the nodes inside the crystal, thereby selecting a different group of phonons to mediate optomechanical coupling (See Supplementary Note: II D for details).
%
%
%
\begin{figure*}[]
\centerline{
\includegraphics[width=12cm]{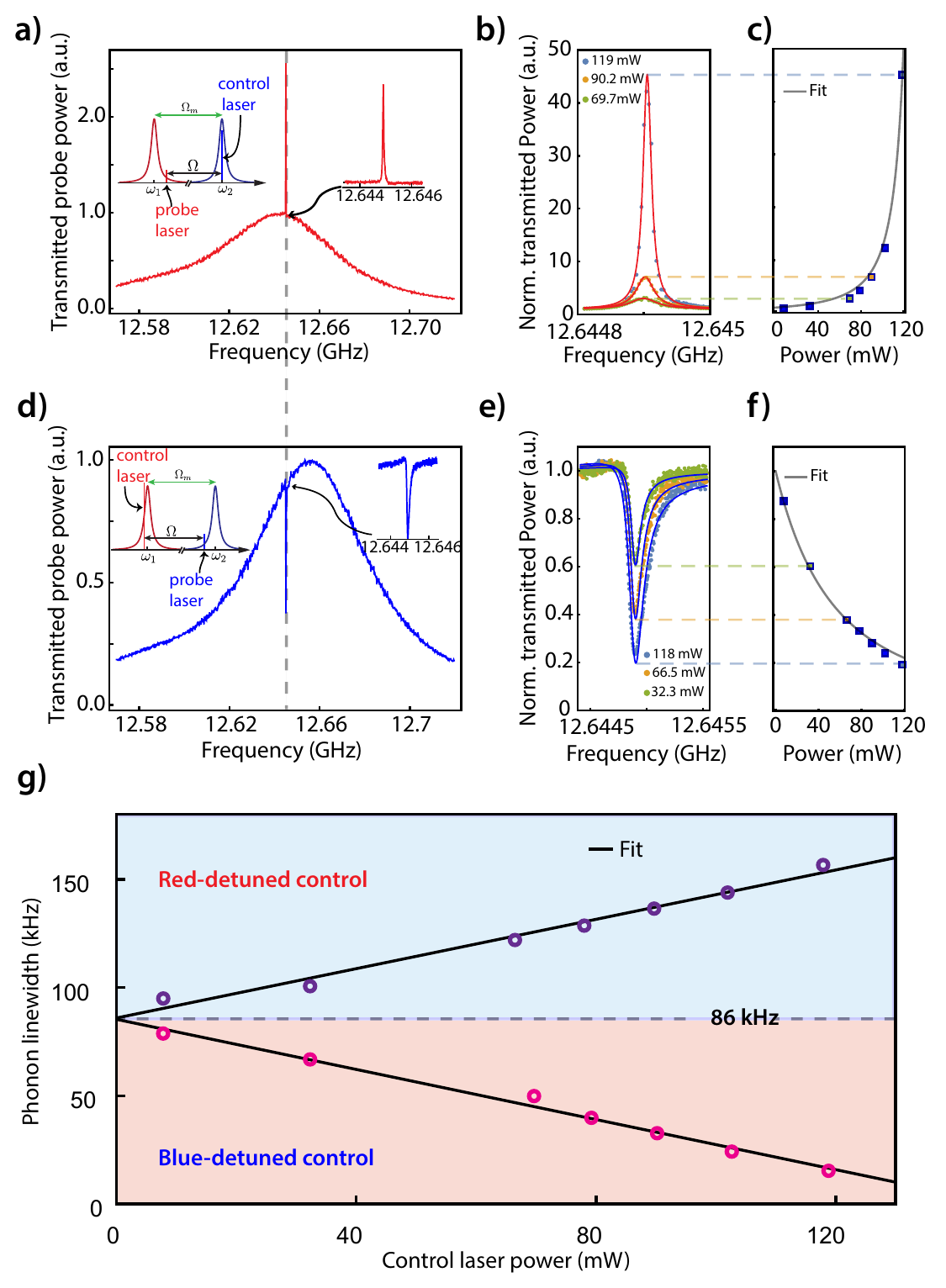}}
\caption{\textbf{Characterization of the zero-point coupling rate and the mechanical dissipation rate.}  
\textbf{a}, We perform OMIA measurements on a single phonon mode using a blue-detuned control laser that is resonant with the high-frequency optical modes (See inset). 
\textbf{b,} As we increase the power in the control laser, the height of the peak corresponding to the OMIA signal increases nonlinearly with control laser power, as displayed in \textbf{c}. 
The linewidth of the OMIA peak, however, decreases with the control power as seen in \textbf{g.}
\textbf{d,} OMIT measurements are performed with the same phonon mode by simply red-detuning the control laser. 
Note that because the control laser has a slight offset relative to the low frequency optical mode at $\omega_j$, the center of the cavity resonance shifts but the OMIT dip is observed exactly at the phonon frequency of $12.645$ GHz. 
\textbf{e}, The dip of the OMIT signal decreases non-linearly as a function of the control power, as displayed in \textbf{f}.
The linewidth of the OMIT signal, however, increases linearly with the control power, as seen in \textbf{g.}
From the dependence of linewidth as a function of the control laser power, we obtain a cold cavity linewidth $\Gamma_m/2\pi \simeq 86$ kHz for the phonon mode at 12.645 GHz.
}
\label{fig4}
\end{figure*}

\section*{Probing coherent optomechanical response}
We explore the optomechanical coupling in our system by probing its coherent response to an optical drive.
%
Specifically, we use the well known techniques called optomechanically induced amplification (OMIA) and optomechanically induced transparency (OMIT)\cite{weis2010optomechanically, dong2015brillouin, kim2015non}. 
%
%
%
%
%

These measurements are performed using a tunable laser at frequency $\omega_l$ whose output is split into two arms (See Fig. \ref{fig3}a). 
Laser light in one arm is intensity modulated at a variable frequency ($\Omega$) using a microwave generator. 
Light in the other arm is frequency-shifted to $\omega_l+2\pi \times 44.0$ MHz using an acousto-optic modulator (AOM).
This AOM-shifted light acts as a local oscillator (LO) such that the Stokes and the anti-Stokes signals appear as distinct tones in the radio-frequency (rf) spectrum analyzer during heterodyne detection. 
The tones at frequencies $\omega_l$ and $\omega_l-\Omega$ serve as control and probe lasers, respectively. 
The strong control laser ($\omega_l$) drives the higher frequency optical mode at $\omega_{j+1}$ whereas the weak probe laser ($\omega_p = \omega_l-\Omega$) is swept near the lower frequency optical mode at $\omega_j$ (See Inset \textbf{i} of Fig. \ref{fig3}a).
The third tone at $\omega_l+\Omega$ is irrelevant, as it is not resonant with the optical cavity modes. 
Light is delivered to and collected from the optical cavity through a combination of fiber collimators and free-space optics.
Light transmitted through the optical cavity is combined with the LO and detected using a photoreceiver, which is connected to a rf spectrum analyzer. 
This spectrum analyzer monitors the beat-note between the transmitted probe laser and the LO by tracking the frequency ($\Omega$) of the microwave generator.
Heterodyne detection of the probe light transmitted through the optical cavity provides a direct measurement of the intra-cavity probe power. 
%


%
This OMIA measurement (Fig. \ref{fig3}b) reveals a broad optical cavity resonance of linewidth $\kappa_j \simeq 2\pi \ \times 73 $ MHz, consistent with the mirrors' reflectivities.
Near the center of the optical resonance, we find three narrow resonances (See inset Fig. \ref{fig3}b) corresponding to phonon modes near the Brillouin frequency of $12.661$ GHz.
%
These resonances are equally spaced by $\sim 612$ kHz, the expected acoustic FSR $v_a/(2L_\text{ac})$.
This result demonstrates optomechanical coupling to multiple high-frequency longitudinal phonon modes, in a manner consistent with the phase-matching described by Eq. (\ref{g0m}).
%

%

In what follows, we tune the optical wavelength to excite a different pair of optical resonances, ensuring that only a single phonon mode mediates the optomechanical interaction. 
A typical OMIA spectrum in which the optomechanical coupling is mediated by a single phonon mode is show in Fig. \ref{fig4}a.
We perform both OMIA and OMIT measurements to determine the associated single-photon coupling rate and the intrinsic mechanical damping rate.
When the system is driven by a strong control laser and a weak probe, we can describe the OMIA (OMIT) phenomena with a linearized interaction Hamiltonian $\hat{H}_\text{int}^{m} = -\hbar g_0^m \sqrt{n_c} (\hat{a}_j \hat{b}_m + \hat{a}_j^\dagger \hat{b}_m^\dagger)$ ($\hat{H}_\text{int}^{m} = -\hbar g_0^m \sqrt{n_c} (\hat{a}_{j+1}^\dagger \hat{b}_m + \hat{b}_m^\dagger \hat{a}_{j+1})$), where $n_c$ is the intra-cavity photon number for the optical mode at $\omega_{j+1} (\omega_j)$. 
Assuming that the control laser is directly on resonance with the optical mode at $\omega_{j+1} (\omega_j)$, this effective Hamiltonian predicts a relative OMIA peak height (OMIT dip) of $1/(1 \mp C)^2$, when $\Omega=\Omega_m$. 
Here, $C = 4 n_c |g_0^m|^2/(\kappa \Gamma_m)$ is the multi-photon co-operativity and $\Gamma_m$ is the instrinsic mechanical damping rate. 
Moreover, the linewidth of this OMIA peak (OMIT dip) is given by $\Gamma_\text{eff} = (1 \mp C)\Gamma_m$  \cite{aspelmeyer2014cavity}.

To measure $\Gamma_m$ and $g_0^m$, we varied the control laser power and measured the relative heights and the linewidths of the OMIA peaks (OMIT dips) (See Fig. \ref{fig4}b,c,g).
%
%
As expected from theory, the relative heights of the OMIA peaks (OMIT dips) increased (decreased) non-linearly whereas the effective linewidth, $\Gamma_\text{eff}$, decreased (increased) linearly as the control laser power was increased from 7.8 mW to 118 mW.
Extrapolating the linear fit in Fig. \ref{fig4}g to zero input power gives $\Gamma_m \simeq 2\pi \times 86$ kHz (acoustic $Q$-factor $\simeq 1.5\times 10^5$).
%
%
The mechanical damping in the present system is dominated by diffractive losses produced by planar acoustic resonator geometry (See Supplementary Note: V). 
These acoustic $Q$-factors can be dramatically increased by shaping the surfaces of the bulk acoustic wave resonator to compensate for the effects of diffraction \cite{renninger2018bulk}. 
The co-operativities ($C$) and intrinsic mechanical damping rate ($\Gamma_m$) obtained from experiments give $g_o^m \approx 2\pi \ \times $ 18 Hz, consistent with the theoretically predicted value of $2\pi \ \times $ 20 Hz (See Supplementary Note: III, IV). 
%
So far, we have probed the system's coherent response. 
Next, we explore thermal fluctuations of the phonon mode and regenerative self-oscillation as we increase the multi-photon co-operativity to greater than unity.
\begin{figure}[]
\centerline{
\includegraphics[width=\linewidth]{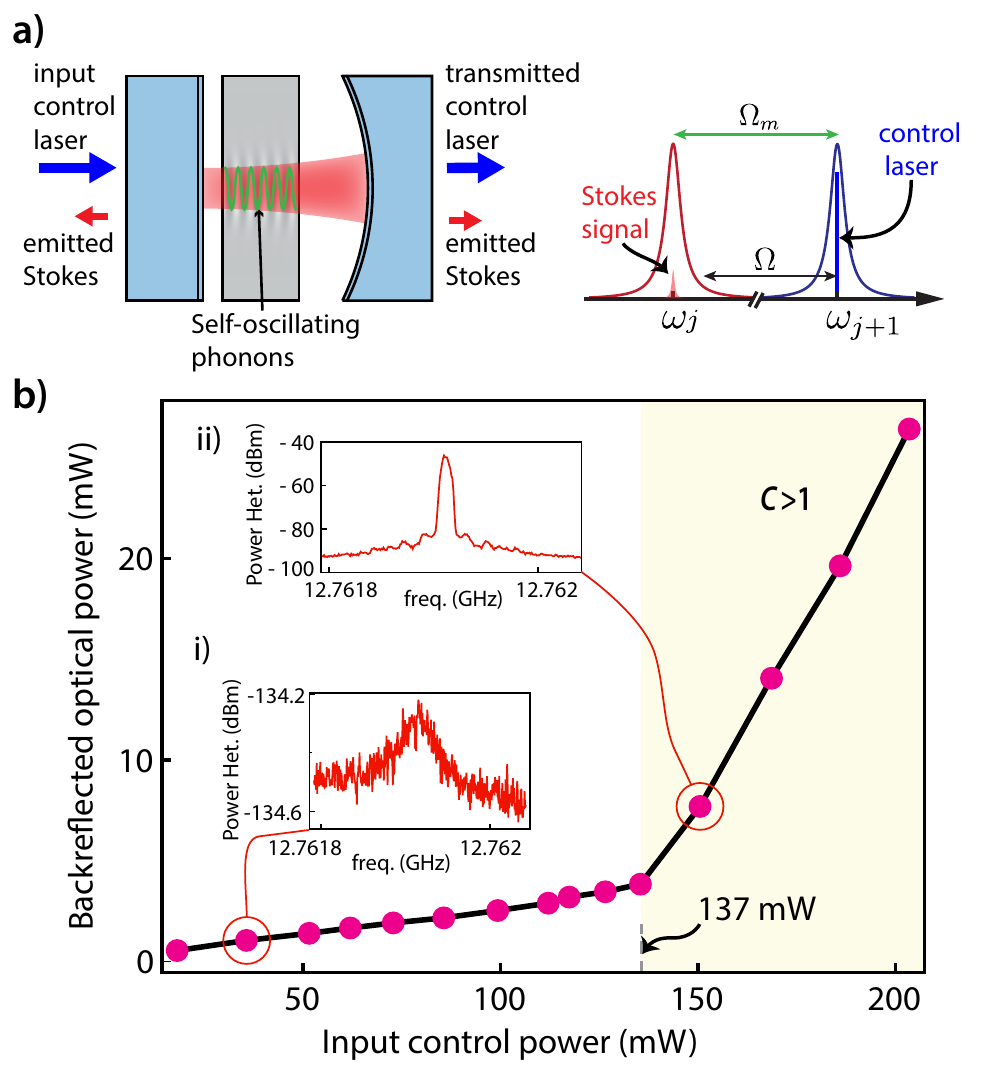}}
\caption{\textbf{Thermal fluctuations and regenerative self-oscillations of a high-frequency phonon mode.}
\textbf{a}, To observe thermal fluctuations of the phonon mode we turn off the probe laser and tune a strong control laser directly on-resonance with the high-frequency optical mode.
The thermally-populated phonon mode spontaneously scatters light from the higher energy optical mode to the lower energy optical mode (Stokes field); this light exits our optical system through both mirrors. 
\textbf{b,} From the measurement of backreflected optical power as a function of the input control laser power, we observe a clear threshold behavior at 137 mW corresponding to regenerative self-oscillation of the phonon mode (i.e. $C>1$). 
Before threshold, a small ($\sim$ 2.6$\%$) backreflection results from imperfect coupling of the control laser to the optical mode at $\omega_2$.
Whereas once self-oscillating, phonons scatter a large fraction of the input control laser into Stokes light, which exits the optical cavity through both forward and backward direction, leading to a significant increase in the backreflected optical power. 
Furthermore, from the heterodyne beat-tone of scattered Stokes light with the frequency shifted version of the input control laser, we observe a significant line narrowing of phonons as we cross the lasing threshold, as displayed in the insets \textbf{i} and \textbf{ii}.
}
\label{fig5}
\end{figure}

\section*{Thermal fluctuations and phonon lasing}
%
We measure thermal fluctuations of the mechanical mode through spontaneous light scattering measurements.
We use a control laser that is resonant with the higher-frequency optical mode $\omega_{j+1}$ (See Fig. \ref{fig5}a);
no probe field is supplied for these measurements.
The thermally populated phonon mode mediates scattering of incident control photons from frequency $\omega_{j+1}$ to $\omega_j$ through Stokes process. 
%
Heterodyne detection is used to monitor the power spectrum of this spontaneously scattered Stokes light as shown in the inset of Fig. \ref{fig5}b-i.
As the control laser power is increased, we observe a sharp increase in the magnitude of the scattered Stokes light accompanied by spectral narrowing of the heterodyne beat tone (inset Fig. \ref{fig5}b-ii).
%
This behavior results from the regenerative self-oscillation of the phonon mode.
The measurement of backscattered optical power from the cavity as a function of the input control laser power (Fig. \ref{fig5}b) reveals a self-oscillation threshold of 137 mW, consistent with the threshold ($\sim$ 140 mW) predicted from the measured values of $g_0^m$, $\kappa$, and $\Gamma_m$. 
The total output Stokes power after lasing is consistent with a slope efficiency of 62 \% (See Supplementary Notes: VI.A for details).
Note that intra-cavity photon number $n_c \approx 6.3 \times 10^9$ is achieved for the highest available control laser power of 204 mW.
Such a large intra-cavity photon number produces cavity-enhanced coupling rate of $g_m=\sqrt{n_c}g_0^m= 2\pi \ \times $ 1.5 MHz and $C= 1.4$.
Note that this coupling rate is already more than 10 times larger than $\Gamma_m$, as required for high-fidelity transduction of quantum information from the optical to the mechanical domain (or vice versa) \cite{zhang2003quantum}.

\section*{Discussion}
These results lay the foundation for a promising new class of macroscopic cavity optomechanical systems that rely on bulk acoustic modes of a crystalline solid---rather than sub-wavelength structural control---to achieve high-frequency multimode interactions. 
Since phase-matching determines the phonon frequency, this approach permits coupling to massive (20 $\upmu$g), high-frequency phonon modes without size reduction.
Resonant optical driving of this multimode system produces appreciable coupling rates (1.5 MHz) to high-frequency (13 GHz) phonons and $C > 1$. 
These results are obtained using the simplest of flat-flat crystal geometries, meaning that these same principles can be readily used to transform practically any transparent crystal into a high-frequency cavity optomechanical system.
Since the Brillouin frequency depends on the optical wavelength and material parameters, these same strategies can be used to harness phonons over a tremendous range of frequencies (e.g., 5-100 GHz) by designing the system around different wavelengths and materials.
The versatility and robustness of this strategy should lend itself to new types of hybrid quantum systems, new forms of materials spectroscopy, and studies of laser-oscillator physics.


Our quartz-based optomechanical system has many promising features in the context of cavity quantum optomechanics.
Interestingly, large optomechanical coupling rates ($> 60$ MHz) to these high-frequency phonons could be achieved even without miniaturization because this macroscopic system can store large number of intra-cavity photons ($>10^{13}$) through resonant driving \cite{meng2005damage}. 
In fact, the demonstrated coupling rate of 1.5 MHz is already large enough to enter the strong-coupling regime ($g_m>\kappa,\ \Gamma_m$) if the optical finesse is boosted from its current value ($\mathcal{F} \approx 170$) to $\mathcal{F} \approx 10^4$.
In this regime, one can deterministically swap excitations between the optical and the phononic domains for quantum transduction and the creation of non-classical mechanical states \cite{aspelmeyer2014cavity}. 
Moreover, because of the potential for reduced thermal decoherence and the opportunity to reach phonon counting sensitivities of less than one, this system shows great promise as a platform to implement probabilistic schemes for quantum state preparation (See Supplementary Notes: VIII for details). 
Owing to the low (high) optical absorption (thermal conductivity) ($<$ 4.3 dB/km \cite{pinnow1973development}) of pristine crystalline quartz and greatly reduced ($<10^{-4}$) photon-surface interactions relative to high-frequency microscale counterparts \cite{eichenfield2009optomechanical}, this system offers a promising path to robust quantum optomechanics.


Hybrid quantum systems utilizing long-lived phonons within bulk acoustic resonators could be a valuable resources for quantum information processing. 
It is possible to increase the $Q$-factor of such high-frequency phonons within quartz from their present values ($1.5\times 10^5$) to $5 \times 10^7$ by shaping the acoustic resonator into a plano-convex geometry \cite{renninger2018bulk}.
These highly coherent bulk acoustic modes could then be useful as quantum memories. 
Moreover, such high-Q phonon modes could readily permit strong coupling to individual defect centers for quantum information processing \cite{soykal2011sound, ruskov2013chip} (See supplementary Note: V for details).
Simultaneous optomechanical and electromechanical control of bulk acoustic wave phonons within piezoelectric crystals also offers a path towards high fidelity microwave-to-optical conversion \cite{bochmann2013nanomechanical, andrews2014bidirectional}.

Beyond the conventional goals of quantum optomechanics, this optomechanical system presents new opportunities for materials spectroscopy and precision measurement. 
Sensitive metrology of cryogenic phonon physics and defect centers can be a performed in a wide array of materials to understand various decoherence channels for phonons.
Moreover, bulk acoustic resonators show great potential for quantum-noise-limited optomechanical oscillators with ultra-narrow fundamental linewidth ($<$ 1 nHz) as they can support large coherent phonon populations ($>10^{12}$) (See Supplementary Note: VI.B for details).
Such highly coherent oscillators could be used for precision sensing \cite{jensen2008atomic, chaste2012nanomechanical}, time keeping \cite{tobar2000cryogenically, hossein2010optomechanical} and the exploration of new physics \cite{wolf2003tests, goryachev2014gravitational}. 

\section*{Methods}
For the theoretical estimate of the coupling rate we use the following parameters: $n = 1.55$, $p_{13}=0.27$, $\rho = 2648$ kg/m$^3$, $L_\text{opt} = 9.13$ mm, $L_\text{ac} = 5.19 $ mm, $A \simeq \pi \times (61 \ \upmu \text{m})^2$, $\omega_j = 2\pi \times 193.4$ THz, $\Omega_m = 2\pi \times 12.65 $ GHz (See Supplementary Notes: II.D for more details). 

The Gaussian optical modes with waist $w_\text{opt} \simeq 61 \ \upmu $m, drive a longitudinal standing wave phonon mode with Gaussian transverse profile with waist $w_\text{ac} \simeq 43 \ \upmu$m. Therefore, the effective motional mass of the phonon mode in our system $m_\text{eff} \approx \pi \rho L_\text{ac} w_\text{ac}^2/4 = 20 \ \upmu$g \cite{pinard1999effective}.

\subsection*{Acknowledgments}
We acknowledge funding support from ONR YIP (N00014-17-1-2514), NSF MRSEC (DMR-1119826),  AFOSR (FA9550-09-1-0484 and FA9550-15-1-0270), ONR MURI on Quantum Optomechanics (Award No. N00014-15-1-2761), and the Packard Fellowship for Science and Engineering. N.T.O. acknowledges support from the National Science Foundation Graduate Research Fellowship under Grant No. DGE1122492. The authors thank Yiwen Chu, Vijay Jain, Shai Gertler, Taekwan Yoon, Charles Brown, Yishu Zhou, and Yizhi Luo for helpful discussions and feedback. The authors of this paper are contributors to patent application no. 62/465101 related to Bulk Crystalline Optomechanics, which was submitted by Yale University.

\subsection*{Author Contributions}
P.K., G.I.H., and E.A.K. performed the experiments and P.K. analyzed the data with input from J.G.E.H and P.T.R. P.K. developed the analytical theory with guidance from G.I.H, J.G.E.H and P.T.R.  W.H.R. contributed to the models of photon-phonon coupling and N.T.O. aided in the development of experimental techniques. All authors participated in the writing of this manuscript.

\subsection*{Competing Interests}
The authors declare no competing financial interests.

\subsection*{Correspondence}
Correspondence and requests for materials should be addressed to P. Kharel (email: prashanta.kharel@yale.edu) or P. T. Rakich (peter.rakich@yale.edu ).

\subsection*{Data Availability}
The data that support the plots within this paper and other findings of this study are available from the corresponding authors upon reasonable request.


\newpage\null\thispagestyle{empty}\newpage



\section*{Supplementary material}

**CONTENTS**

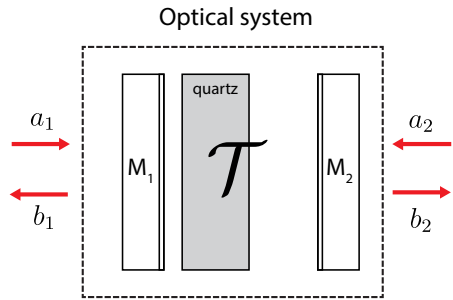

**Optical system**

FIG. 1. **Scattering Matrix Treatment**. A cartoon of our optical system, which consists of two optical mirrors with a quartz crystal placed in between them. A transfer matrix ($\mathcal{T}$) relates the incoming and outgoing optical fields in our system.

# I. ASYMMETRIC CAVITY MODE SPACING

In this section, we use scattering/transmission matrix approach to calculate the mode spectrum of an optical cavity consisting of a bulk crystal placed in between two mirrors. Unlike equally spaced standing-wave longitudinal optical modes of an optical cavity in vacuum, we show that our optical system supports a rich mode spectrum with mode spacings that vary dramatically as a function of both the optical mode number and the crystal position.

Our one dimensional model consists of a Fabry-Perot optical cavity having power reflectivities of $R_1$ and $R_2$ and transmission of $T_1$ and $T_2$. A dielectric material of index $n$ placed in between the two mirrors. We assume the mirrors are lossless so that $R_i + T_i = 1$, where $i = 1, 2$. The power reflectivity at each crystal face, $R_0$, is given by $(1-n)^2/(1+n)^2$. For simplicity, we assume plane wave optical fields and focus on the cavity's reflection and transmission spectrum.

We define the plane wave optical fields propagating along positive direction as $E = \text{Re}\{ae^{i(kz-\omega t)}\}$. The phase $\phi$ acquired by the plane wave after propagating a distance $z$ in a medium with refractive index $n$ is given by $\phi = kz = n\omega z/c$, where $c$ is the speed of light in vacuum. A transmission matrix relates the input field amplitudes $a_1$, $a_2$ and output field amplitudes $b_1$, $b_2$ (See Fig. 1) in our system as follows:

$$\begin{bmatrix} b_1 \\ a_1 \end{bmatrix} = \mathcal{T} \begin{bmatrix} b_2 \\ a_2 \end{bmatrix} \tag{1}$$

The transmission matrices for a lossless mirror and for propagation in space are given by [1, 2]

$$\mathcal{T}_{\text{mirror}} = \frac{-i}{t} \begin{bmatrix} -1 & r \\ -r & 1 \end{bmatrix} \text{ and } \mathcal{T}_{\text{prop}} = \begin{bmatrix} e^{i\phi} & 0 \\ 0 & e^{-i\phi} \end{bmatrix}. \tag{2}$$

The total transmission matrix for our optical system consisting of the bulk crystal inside the Fabry-Perot optical cavity is then given by:

$$\begin{aligned}
\mathcal{T} &= \mathcal{T}_{mirror1} \cdot \mathcal{T}_{propVacuum} \cdot \mathcal{T}_{mirrorQuartz} \cdot \mathcal{T}_{propQuartz} \cdot \mathcal{T}_{mirrorQuartz} \cdot \mathcal{T}_{propVacuum} \cdot \mathcal{T}_{mirror2} \\
&= \frac{-i}{t_1} \begin{bmatrix} -1 & r_1 \\ -r_1 & 1 \end{bmatrix} \cdot \begin{bmatrix} e^{i\phi_1} & 0 \\ 0 & e^{-i\phi_1} \end{bmatrix} \cdot \frac{-i}{t_0} \begin{bmatrix} -1 & r_0 \\ -r_0 & 1 \end{bmatrix} \cdot \begin{bmatrix} e^{i\phi_2} & 0 \\ 0 & e^{-i\phi_2} \end{bmatrix} \cdot \frac{-i}{t_0} \begin{bmatrix} -1 & -r_0 \\ r_0 & 1 \end{bmatrix} \\
&\cdot \begin{bmatrix} e^{i\phi_3} & 0 \\ 0 & e^{-i\phi_3} \end{bmatrix} \cdot \frac{-i}{t_2} \begin{bmatrix} -1 & r_2 \\ -r_2 & 1 \end{bmatrix}.
\end{aligned} \tag{3}$$

Here $r_1 = \sqrt{R_1}, r_2 = \sqrt{R_2}, r_0 = (1-n)/(1+n), \phi_1 = \omega t/c, \phi_2 = n\omega L_{ac}/c$, and $\phi_3 = \omega(L_{opt} - L_{ac} - t)/c$. The optical power reflected ($P_r$) from and transmitted ($P_t$) through our optical system can then be calculated from the total transmission matrix as follows

$$P_r = \left| \frac{\mathcal{T}_{12}}{\mathcal{T}_{22}} \right|^2 \text{ and } P_t = \left| \frac{1}{\mathcal{T}_{22}} \right|^2. \tag{4}$$

The analytical expressions for $P_r$ and $P_t$ can be derived in a straightforward way from eqn. (3); we chose to not show them here because they are quite cumbersome to display.

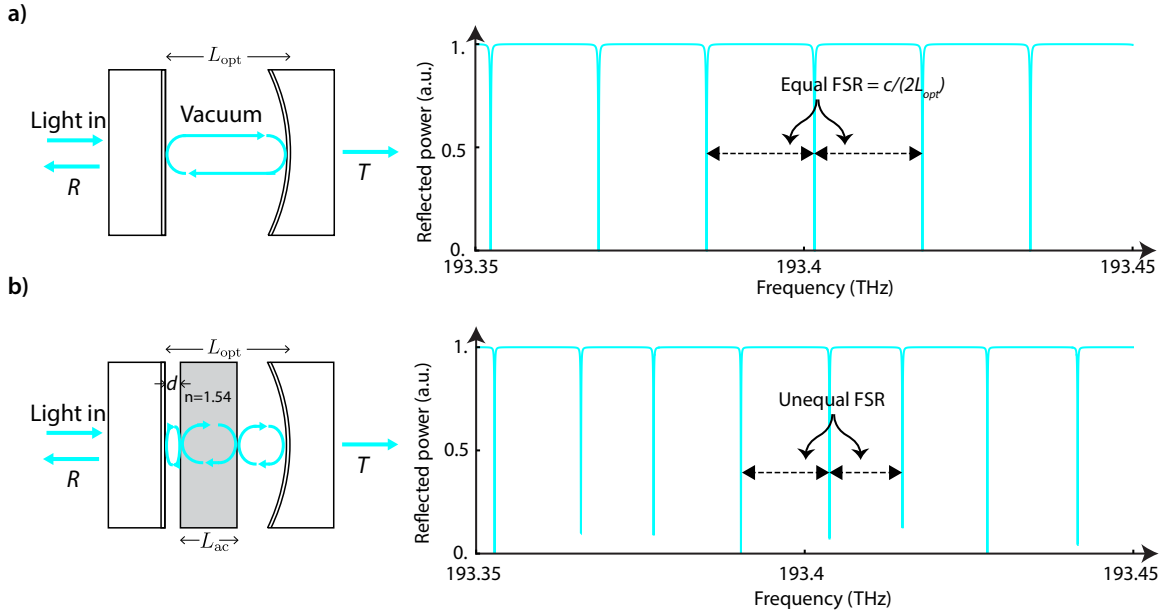

FIG. 2. **Comparison of the mode spectrum of an optical cavity with and without the bulk crystal.** **a,** For an optical cavity in vacuum, the standing wave cavity modes are spaced equally by the optical free spectral range (FSR) given by $c/2L_{opt}$. **b,** The mode spectrum of an optical cavity quite dramatically when a medium with refractive index $n$ is placed in between the two mirrors. As expected the frequency spacing between the optical modes decreases because of increase in the total optical path length. In addition, there is a variation in the optical free spectral range (i.e., change in the spacing between adjacent optical modes) as a function of frequency. Surprisingly, a large variation in FSR ($\sim 21\%$) occurs even through the reflectivity of the quartz-vacuum interface is rather small ($\sim 4\%$).

| Parameters | values |
|------------|--------|
| $R_1$ | 0.98 |
| $R_2$ | 0.98 |
| $d$ | 0.15 mm |
| $L_{ac}$ | 5.19 mm |
| $L_{opt}$ | 9.13 mm |
| $n$ | 1.55 |

TABLE I. Table 1: Parameters relevant to our optical cavity system, which consists of a quartz crystal placed in between the mirrors.

Next, we use the analytical expression for $P_r$ along with the known geometrical parameters (see Table I) and mirror reflectivities to explore the reflection spectrum of our optical cavity (See Fig. 2). We compare the cavity mode spectrum with and without the quartz crystal inside the optical cavity near experimentally relevant wavelength (1550 nm or $\omega/2\pi \sim 194$ THz). Without the crystal inside the optical cavity, this analytical calculation reveals equally spaced cavity modes, which are separated by the well known optical free spectral range given by $c/(2L_{opt})$. However, when the quartz crystal is placed inside an optical cavity, we notice that the average spacing between the optical modes gets narrower; this result makes sense as the optical path length is longer when the crystal is placed inside the cavity. More notably, we observe a variation in free spectral range (i.e., change in the spacing between adjacent optical modes) as a function of cavity frequency.

We explored the variation in FSR analytically by changing the parameter $d$, corresponding to the position of the crystal inside the cavity (See Fig 3a). We observed periodic variations in cavity mode frequencies for each mode number $j$ (See Fig. 3b) as a function of the crystal displacement ($\Delta d$). This variation in cavity frequency was $\lambda/2$ periodic, where $\lambda$ is the wavelength of the light; this is a result of the crystal surfaces passing through the nodes and anti-nodes of the standing wave optical cavity modes. Notice that the amount of frequency variation is different for each cavity mode. Consequently, the free-spectral range changes both as a function of the cavity mode number ($j$) and the crystal position. Therefore, it is relatively easy to tune the FSR to match the Brillouin frequency in this system.

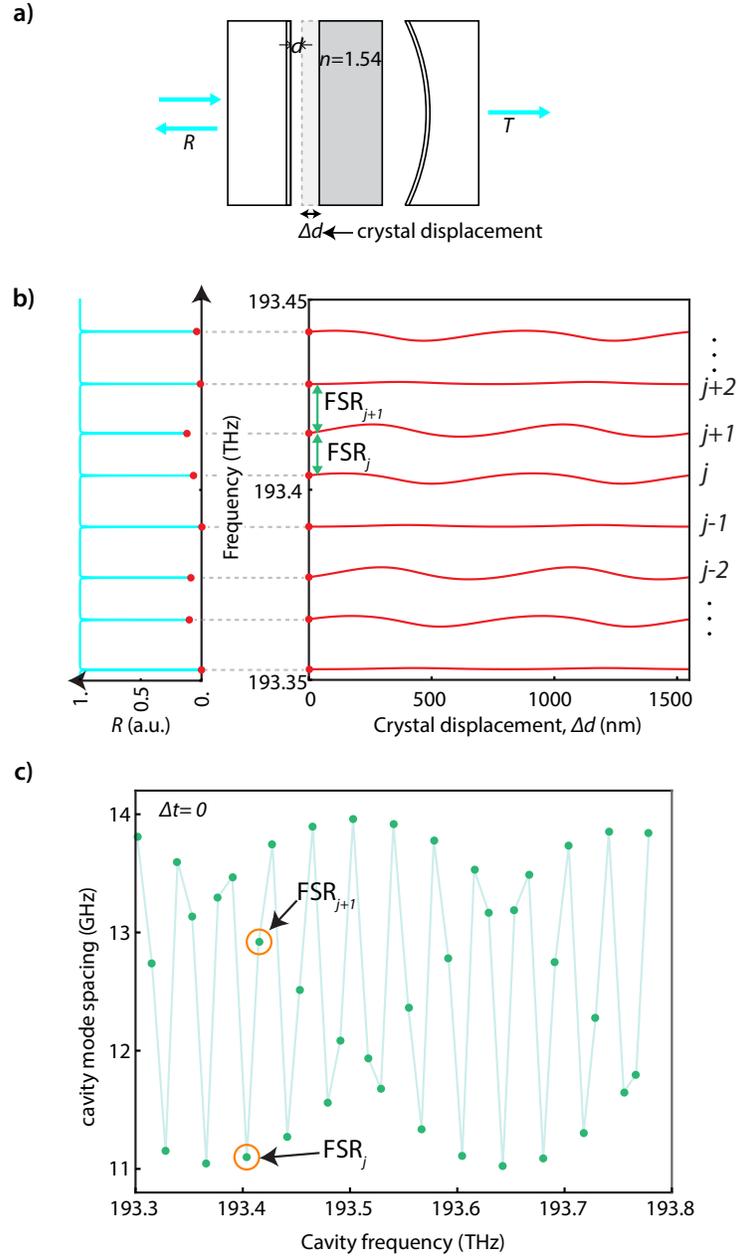

FIG. 3. **Variation in optical cavity frequency as a function of the position of the crystal inside the cavity. a**, A schematic showing how the crystal is displaced from its original location $d$ to $d + \Delta d$. **b**, Plot of the resonant cavity modes, labeled with the longitudinal mode number $j$, reveals periodic variation in the mode frequency as a function of the cavity displacement ($\Delta t$). This periodicity is equal to half the wavelength of light ($\sim$1550 nm), corresponding to the crystal moving through the nodes and the anti-nodes of the standing wave optical cavity modes. Although periodic, the extent of variation in the cavity frequency is different for each mode; as a result, we obtain a non-trivial variation in the cavity FSR as a function of optical mode number $j$ seen in **c**.

From theory, we expect a maximum FSR variation as large as 2.94 GHz between 1548 nm and 1552 nm (See Fig. 3c). Furthermore, the difference in FSR between adjacent optical modes ($\chi_j = \text{FSR}_{j+1} - \text{FSR}_j$) can be as large as 2.47 GHz; the parameter $\chi_j$ is important because it quantifies the degree of asymmetry between the Stokes and the anti-Stokes scattering processes. Additionally, since the FSR variation is periodic as a function of cavity frequency, it is possible to find several pairs of cavity modes having FSR that equal Brillouin frequency.

Next, we compare experimental measurement of the cavity mode spacing and compare it with theory. The reflection spectrum of the cavity was measured at cryogenic temperature by sweeping the laser wavelength between 1548 nm

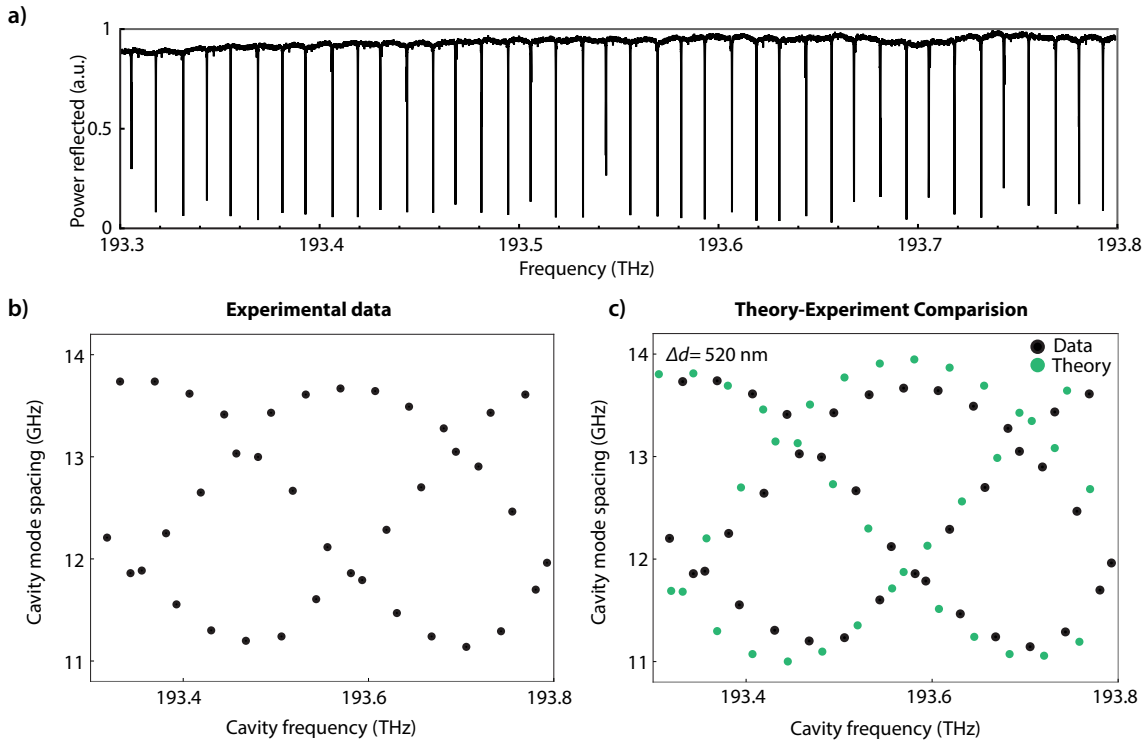

FIG. 4. **Measurement of the power reflected from the optical cavity as a function of the laser frequency at cryogenic temperature ($\sim$ 8K). a,** Measurement from 1548 nm to 1552 nm (or 193.3 GHz to 193.8 GHz) revealed narrow optical resonances corresponding to longitudinal cavity modes. Some modes have larger dip depths than others as expected from theoretical calculations. **b,** We see periodic variation in FSR as a function of the cavity frequency (or mode number $j$). This variation is very similar to what was predicted from theoretical analysis. **c,** To compare theory with experiments, we first change the crystal displacement ($\Delta d$); we take crystal displacement as a fit parameter as the relative spacing between the crystal and the mirror can change during the cooldown process. When we choose $\Delta d = 520$ nm, we see that the theoretically obtained mode spacings matches well with the experimentally obtained data. The maximum FSR variation of 2.6 GHz observed experimentally is close to the theoretically predicted value of 2.94 GHz.

and 1552 nm and recording the back-reflected power (See Fig. 4). We first identify the resonant frequencies, $\omega_j/2\pi$, from this measurement. We then calculate the cavity mode spacing by simply taking the difference of two adjacent mode frequencies (i.e., FSR$_j = (\omega_{j+1} - \omega_j)/2\pi$). This analysis revealed periodic variation (See Fig. 4b) in cavity mode spacing, which was similar to the one predicted from theoretical calculation as seen in Fig. 3c. However, because of the uncertainties in the geometrical parameters (such as a exact crystal position $t$) as we cool our system to cryogenic temperatures, we do not precisely know the optical mode numbers $j$. Therefore, we took the exact crystal location as a free parameter. We found from our analytical calculation that changing the crystal location seemed to essentially shift the phase and increase the periodicity in the cavity FSR as a function of the optical cavity frequency. For $t = 152.52$ μm, we see a good agreement between the experimentally determined cavity mode spacing (black dots) and the theoretically calculated values (green dots) in Fig. 4c. With the quartz crystal inside the optical cavity, the maximum FSR variation of 2.6 GHz obtained experimentally agrees well with the predicted value of 2.94 GHz.

## II. HAMILTONIAN TREATMENT

In this section, we derive the total Hamiltonian for our opto-mechanical system. For this we derive the Hamiltonian for the optical fields, the Hamiltonian for the acoustic fields, and finally add an interaction term that characterizes the acousto-optic interaction.

### A. Hamiltonian for the electromagnetic fields

We consider the case where electromagnetic field has a single single polarization ($\hat{x}$−direction) and is subjected to the boundary conditions defined by the two mirrors. We expand the electric field into the normal modes (standing waves) of the optical cavity [3]

$$E_x(z,t) = \sum_j E_j \sin(k_j z)(\hat{a}_j(t) + \hat{a}_j^\dagger(t)),$$  (5)

where $\hat{a}_j$ is the normal mode amplitude, $k_j = j\pi/L_{\text{opt}}$, with j = 1, 2, 3, ..., and the zero point amplitude of the electric field is

$$E_j = \sqrt{\frac{\hbar\omega_j}{\epsilon_o\epsilon_r A_{\text{opt}} L_{\text{opt}}}},$$  (6)

where $\omega_j$ is the frequency of the normal mode, $\epsilon_r$ is the relative permittivity of the optical cavity, $A_{\text{opt}}$ is the transverse area of the optical mode, and $L_{\text{opt}}$ is the optical cavity length. The zero point fluctuation of the electric field was obtained knowing that the total electromagnetic energy per mode for the ground state is $\hbar\omega_j/2$. Starting with the Hamiltonian for the electromagnetic fields

$$H^{\text{opt}} = \frac{1}{2}\int_V dV \ (\epsilon E^2 + \mu_o H^2)$$  (7)

and substituting the normal mode expansions for the electromagnetic fields, we obtain the quantized version of the electromagnetic Hamiltonian given by

$$\hat{H}^{\text{opt}} = \sum_j \hbar\omega_j \left(\hat{a}_j^\dagger \hat{a}_j + \frac{1}{2}\right).$$  (8)

Note that the mode amplitude operators satisfy the commutation relations $[\hat{a}_j, \hat{a}_{j'}^\dagger] = \delta_{j,j'}$ and $[\hat{a}_j, \hat{a}_{j'}'] = [\hat{a}_j^\dagger, \hat{a}_{j'}^\dagger] = 0$.

### B. Hamiltonian for the acoustic fields

For the quantization of acoustic fields, we consider a longitudinally polarized ($\hat{z}$-direction) acoustic field subject to the free boundary conditions at the surfaces of the crystal. We expand the acoustic displacement field into the normal modes (standing waves) of the Fabry-Perot acoustic cavity

$$u_z(z,t) = \sum_m U_m \cos(q_m z)(\hat{b}_m(t) + \hat{b}_m^\dagger(t))$$  (9)

where $\hat{b}_m$ is the normal mode amplitude, $q_m = m\pi/L_{ac}$, with $m = 1, 2, 3, ...$, and the zero point amplitude of the acoustic displacement field

$$U_m = \sqrt{\frac{\hbar}{\rho A_{\text{ac}} L_{\text{ac}} \Omega_m}},$$  (10)

where $\Omega_m$ is the frequency of the acoustic mode, $\rho$ is the density of the medium, $A_{\text{ac}}$ is the transverse area of the acoustic mode, and $L_{\text{ac}}$ is the thickness of the acoustic Fabry-Pérot cavity. The zero point fluctuation of the acoustic field was calculated knowing that the total acoustic energy (kinetic plus potential) per mode for the ground state is $\hbar\Omega_m/2$. We start with the acoustic Hamiltonian [4, 5]

$$H^{\text{ph}} = \frac{1}{2}\int_V dV \ \left(\rho\dot{u}^2 + C\left(\frac{\partial u}{\partial z}\right)^2\right),$$  (11)

where $C = v_{\text{ac}}^2 \rho$ is the elastic coefficient and substitute the normal mode expansion to obtain the quantized version of the acoustic field

$$\hat{H}^{\text{ph}} = \sum_m \hbar\Omega_m \left(\hat{b}_m^\dagger \hat{b}_m + \frac{1}{2}\right).$$  (12)

Note that the mode amplitude operators satisfy the commutation relations $[\hat{b}_m, \hat{b}_{m'}^\dagger] = \delta_{m,m'}$ and $[\hat{b}_m, \hat{b}_m'] = [\hat{b}_m^\dagger, \hat{b}_{m'}^\dagger] = 0$.

## C. Interaction Hamiltonian

We now consider the interaction Hamiltonian for the optomechanical coupling [5], which is given by

$$H^{\text{int}} = \frac{1}{2} \int_V \, dV \; \epsilon_o \epsilon_r^2 E^2 p_{13} \frac{\partial u}{\partial z}, \tag{13}$$

where $p_{13}$ is the relevant photoealstic constant. Substituting the normal mode expansions for the electric field and the acoustic displacement field from eqn. (5) and eqn. (9) into this interaction Hamiltonian and using the rotating wave approximation, we obtain

$$\hat{H}^{\text{int}} = - \sum_{j,j',m} \int \, dV \; \epsilon_o \epsilon_r^2 p_{13} q_m U_m E_j E_{j'} \, \sin(k_j z)\sin(k_{j'} z)\sin(q_m z) \; (\hat{a}_j^\dagger \hat{a}_{j'} \hat{b}_m + \hat{a}_j \hat{a}_{j'}^\dagger \hat{b}_m^\dagger) \tag{14}$$

The term $\hat{a}_j^\dagger \hat{a}_{j'} \hat{b}_m$ in the interaction Hamiltonian represents the annihilation of an optical mode at frequency $\omega_{j'}$ and a phonon mode at frequency $\Omega_m$ to create a photon mode at higher frequency $\omega_j = \omega_{j'} + \Omega_m$ (also called the anti-Stokes process). The other term $\hat{a}_j \hat{a}_{j'}^\dagger \hat{b}_m^\dagger$ represents the conjugate process, whereby a photon at lower frequency $\omega_{j'}$ and a phonon at frequency $\Omega_m$ is created from an annihilation of photon at $\omega_j$ (Stokes process).

Now, if we consider two adjacent standing wave optical modes such that $\omega_{j+1} - \omega_j = \Omega_m$, the interaction Hamiltonian for the coupling to the $m$-th phonon mode is given by

$$\hat{H}_m^{\text{int}} = -\hbar g_0^m(\hat{a}_{j+1}^\dagger \hat{a}_j \hat{b}_m + \text{H.c.}). \tag{15}$$

where the single-photon coupling rate for this multi-mode optomechanical system is given by

$$g_0^m = \frac{1}{\hbar} \int \, dV \; \epsilon_o \epsilon_r^2 p_{13} q_m U_m E_{j+1} E_j \, \sin(k_{j+1}z)\sin(k_j z)\sin(q_m z). \tag{16}$$

## D. Single-Photon Coupling Rate

We use the definition of the coupling rate given in eqn. (16) along with the zero-point amplitudes of the electric and the acoustic fields (eqn. (6) and eqn. (10)) to calculate the single photon coupling rate,

$$g_0^m = \frac{1}{\hbar} \int \, dV \; \epsilon_o \epsilon_r^2 p_{13} q_m U_m E_{j+1} E_j \, \sin(k_{j+1}z) \, \sin(k_j z) \, \sin(q_m z)$$

$$= A\epsilon_o \epsilon_r^2 p_{13} q_m \sqrt{\frac{\hbar}{\rho A L_{\text{ac}} \Omega_m}} \frac{\sqrt{\omega_j \omega_{j+1}}}{\epsilon_o \epsilon_r A L_{\text{opt}}} \times$$

$$\int_d^{d+L_{\text{ac}}} dz \, \sin\left(k_{j+1}\left(z - d + \frac{d}{n}\right)\right) \sin\left(k_j \left(z - d + \frac{d}{n}\right)\right) \sin\left(q_m(z - d)\right) \tag{17}$$

where $d$ is a variable crystal position, and we are considering only the optical fields living inside the crystal for the acousto-optical overlap in $z$. The phase factors for the sine functions come from the appropriate boundary conditions on the electric and acoustic fields.

Using a simple trigonometric identity,

$$\sin(A) \cdot \sin(B) \cdot \sin(C) = 1/4(-\sin(A-B-C)+\sin(A+B-C)+\sin(A-B+C)-\sin(A+B+C)), \tag{18}$$

we see that non-zero spatial overlap and, hence, a non-zero coupling rate occurs when the phase matching requirement ($q_m = k_{j+1} + k_j$) is satisfied. Note that this equation also shows that one can change the coupling rate by changing the crystal position. For example, as the crystal position is changed along the $z$-direction the nodes of the strain profile could line up with the anti-nodes of the optical beat tone (or the forcing function) resulting in zero optomechanical coupling. For maximum coupling rate to a single acoustic mode, the crystal position (or the optical wavevector) has

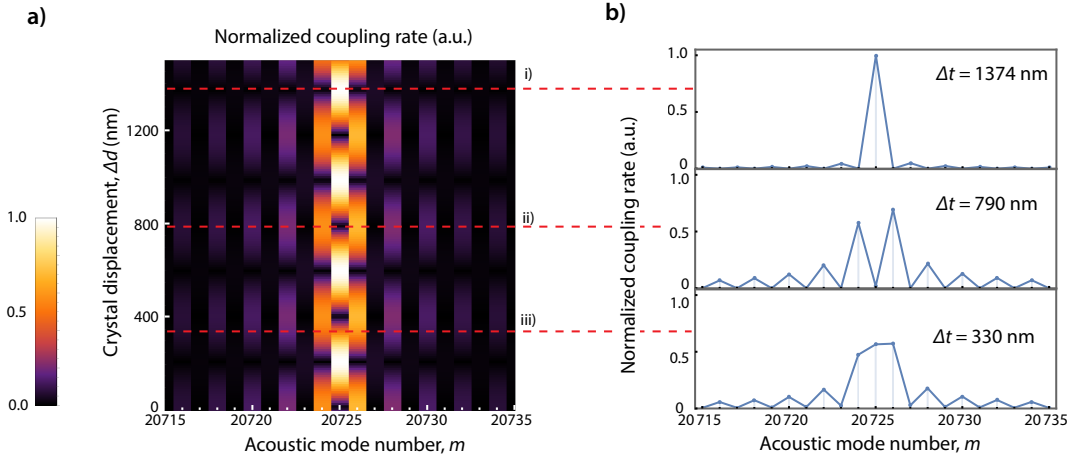

FIG. 5. **Normalized coupling rate** ($|g_0^m|/\tilde{g}_0$) **as a function of acoustic mode number,** $m$ **and the crystal displacement,** $\Delta d$. **a**, We observe that appreciable optomechanical coupling occurs over a small set of phonon modes (indexed by longitudinal mode number $m$) near the Brillouin frequency. The optomechanical coupling to a single mode (say m = 20725) is periodic and can go to zero at certain crystal positions. Nevertheless, for an arbitrary crystal displacement, we still have an appreciable coupling at least one phonon mode. **b**, Line cuts in the 2-Dimensional density plot of the normalized coupling rate shows that tailorable coupling to one (inset **i**), two (inset **ii**) or three phonon modes (inset **iii**) can be achieved by changing the crystal displacement.

to be such that the nodes (anti-nodes) of the strain profile line up with the nodes (anti-nodes) of the optical beat tone. In this case, the maximum coupling rate can be calculated from eqn. (17) and is given by

$$\tilde{g}_0 \approx \frac{\omega_j^2 n^3 p_{13}}{2c} \sqrt{\frac{\hbar}{\rho A L_{ac} \Omega_m}} \frac{L_{ac}}{L_{opt}},$$

where we assumed $\omega_{j+1} \simeq \omega_j$, and $q_m = k_{j+1} + k_j \approx 2n\omega_j/c$. We note that the phase matching ($q_m = k_{j+1} + k_j$) and energy conservation ($\omega_{j+1} - \omega_j = \Omega_m$) is satisfied for modes near $\Omega_m \approx 2\omega_j n v_a/c$, where $v_a$ is the velocity of sound inside the medium.

For instance, for two optical modes near 1551.0335 nm (i.e., $\omega_j = 2\pi \times 193.4$ THz) separated by the phonon frequency of $\Omega_m/2\pi = 12.645$, GHz we calculate a maximum coupling rate $g_0^m/2\pi \simeq 20$ Hz in our system. Note that we used the following experimentally relevant material and geometric parameters for the calculation of $g_0^m$: $p_{13} = 0.27$, $n = 1.55$, $L_{ac} = 5.19$ mm, $L_{opt} = 9.13$ mm, and $A = \pi \times (61 \ \mu m)^2$. Note that the transverse mode area, $A$, for resonant optical mode can be calculated simply using ABCD matrix for Gaussian beam propagation [1].

Next, we explore the coupling rate variation as a function of the crystal position for several longitudinal acoustic modes near the Brillouin frequency. We define a normalized coupling rate as $|g_0^m|/\tilde{g}_0$, and calculate optomechanical coupling to phonon modes the Brillouin frequency; the Brillouin frequency of 12.645 GHz corresponds to a very large mode number (m= 20725). It is important to note that the precise value of the mode number is not well known due to the uncertainties in the geometrical parameters and elastic constants at cryogenic temperatures. Our aim here is to simply compare optomechanical coupling to the longitudinal modes near the Brillouin frequency. A density plot in Fig. 5a reveals that appreciable optomechanical coupling occurs to a few modes near the Brillouin frequency. This occurs because of the finite spatial integral in eqn. (17); as a result of the finite length of the crystal, there is an uncertainty in the acoustic wave-vector which relaxes the phase matching condition and grants optomecahnical coupling to more than one phonon modes [6]. The bandwidth of coupling ($\Delta\nu_B = 1.76v_a/2L_{ac}$) is $\sim 1.1$ MHz in our system.

The coupling to an individual phonon mode (say m= 20725) is periodic in the crystal displacement (See Fig. 5a) and it can be zero at certain crystal positions (See Fig. 5b.ii). Nevertheless, appreciable optomechanical coupling (normalized coupling rate > 0.5) occurs for at least one phonon mode irrespective of the crystal displacement. Therefore, as long as the linewidth of the optical mode ($\kappa/2\pi$) is greater than the coupling bandwidth ($\Delta\nu_B$), we observe appreciable optomechanical coupling to at least one phonon mode irrespective of the crystal position. For a given pair of optical modes, changing the crystal position using a piezo-actuator should enable controllable optomechanical coupling to one or more phonon modes in our system (See Fig. 5b.i-iii). Alternatively, coupling to one or more phonon modes can be engineered by changing the wavevector of the optical fields by moving to a different pair of

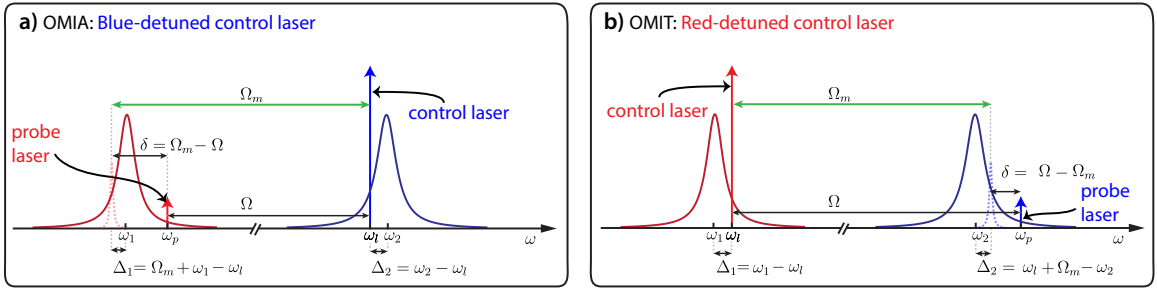

FIG. 6. **Cartoons depicting OMIA and OMIT measurements. a,** A strong blue-detuned control laser is resonant with the high-frequency optical mode while a weak probe is swept near the low frequency optical mode. **b,** A strong red-detuned control laser is resonant with the low frequency optical mode while a weak probe is swept near the high-frequency optical mode.

| Parameters | values |
| --- | --- |
| $\omega_2$ | $2\pi \times 193.419$ THz |
| $\Omega_m$ | $2\pi \times 12.645$ GHz |
| $\kappa_1$ | $2\pi \times 71$ MHz |
| $\kappa_2$ | $2\pi \times 81$ MHz |
| $\kappa_1^{\mathrm{ext}}$ | $\kappa_1/2$ |
| $\kappa_2^{\mathrm{ext}}$ | $\kappa_2/2$ |

TABLE II. Table 2: We choose a pair of optical modes separated by the Brillouin frequency for OMIA/OMIT measurements. The loss rate of each cavity mode was obtained separately from the OMIA/OMIT measurements by sweeping the tunable laser source through the optical resonances and measuring the back-reflected optical power.

optical resonances (with frequency difference matching the Brillouin frequency). We can use eqn. (17) to see that the relative coupling rate to phonon mode near Brillouin frequency changes as we change the wave-vector (or the frequency) of the optical modes modes. This approach avoids the experimental complexity of adding a piezo-actuator in our optomechanical system.

## III. OPTOMECHANCIALLY INDUCED AMPLIFICATION (OMIA)

In this section, we follow the approach outline in Ref. [7] to derive the intra-cavity optical spectrum during OMIA measurements. The optomechanical Hamiltonian accounting for the external drive fields (i.e., the control laser and the probe laser) is given by

$$
\begin{aligned}
H = {} & \hbar\omega_1 a_1^\dagger \hat{a}_1 + \hbar\omega_2 a_2^\dagger a_2 + \hbar\Omega_m b_m^\dagger b_m - \hbar g_0^m (a_2^\dagger a_1 b_m + b_m^\dagger a_1^\dagger a_2) \\
& + i\hbar\sqrt{\kappa_1^{\mathrm{ext}}}\alpha_p (a_1^\dagger e^{-i\omega_p t} - a_1 e^{i\omega_p t}) + i\hbar\sqrt{\kappa_1^{\mathrm{ext}}}\alpha_l (a_2^\dagger e^{-i\omega_l t} - a_2 e^{i\omega_l t}).
\end{aligned}
\tag{19}
$$

We normalize the optical power launched into the cavity such that $P_{in} = \hbar\omega\langle\alpha^\dagger\alpha\rangle$. We assume a strong control laser and a weak probe; within the undepleted pump approximation, the dynamics of the mode at frequency $\omega_2$ is not influenced by the optomechanical coupling and it can be described by the following equation of motion derived from the Hamiltonian in eqn. (19)

$$
\dot{a}_2(t) = \left(-i\omega_2 - \frac{\kappa_2}{2}\right)a_2 + \sqrt{\kappa_2^{\mathrm{ext}}}\alpha_l e^{-i\omega_l t},
\tag{20}
$$

where $\kappa_2 = 2\kappa_2^{\mathrm{ext}} + \kappa_2^0$, $\kappa_2^{\mathrm{ext}}$ is the loss rate at each cavity mirror and $k_2^0$ is the loss rate inside the cavity. We assume negligible internal losses in our system ($k_2^0 \ll 2\kappa_2^{\mathrm{ext}}$). From eqn. (20), we obtain the following steady state solution for mode $a_2$

$$
\langle a_2 \rangle = \sqrt{N_2}e^{-i\omega_l t} = \left| \frac{\sqrt{\kappa_2^{\mathrm{ext}}}\alpha_l}{i\Delta_2 + \kappa_2/2} \right| e^{-i\omega_l t},
\tag{21}
$$

where $N_2$ is the control laser driven intra-cavity photon number for mode $a_2$, and $\Delta_2 = \omega_2 - \omega_l$ (See Supplementary Note Fig. 6). Once we know the intra-cavity photon number for mode $a_2$, we use input-output formalism for a

symmetric Fabry-Perot cavity to obtain the transmitted light field

$$a_{2,\text{out}} = -\sqrt{\kappa_2^{\text{ext}}} a_2. \tag{22}$$

In the undepleted pump regime, we substitute $\langle a_2 \rangle$ from eqn. (21) into the Hamiltonian of eqn. (36) to derive the following linearized Hamiltonian for the interaction between the optical mode $a_1$ and the phonon mode $b_m$

$$H = \hbar\omega_1 a_1^\dagger a_1 + \hbar\Omega_m b_m^\dagger b_m - \hbar g_0^m \sqrt{N_2}(a_1 b_m e^{i\omega_l t} + a_1^\dagger b_m^\dagger e^{-i\omega_l t}) + i\hbar\sqrt{\kappa_1^{\text{ext}}}\alpha_p(a_1^\dagger e^{-i\omega_p t} - a_1 e^{i\omega_p t}). \tag{23}$$

From this Hamiltonian, by rotating in the frame of with $H_0 = \hbar(\omega_l - \omega_p)b_m^\dagger b_m + \hbar\omega_p a_1^\dagger a_1$, we obtain an effective Hamiltonian

$$H_{\text{eff}} = \hbar\delta b_m^\dagger b_m - \hbar(\delta - \Delta_1)a_1^\dagger a_1 - \hbar g_m(a_1 b_m + a_1^\dagger b_m^\dagger) + i\hbar\sqrt{\kappa_1^{\text{ext}}}\alpha_p(a_1^\dagger - a_1), \tag{24}$$

where $\delta = \Omega_m - \Omega = \Omega_m - \omega_l + \omega_p$, and $\Delta_1 = \Omega_m + \omega_1 - \omega_l$, and the cavity field-enhanced coupling rate $g_m = g_0^m \sqrt{N_2}$. The Heisenberg equations of motions derived from this Hamiltonian given by

$$\dot{a}_1(t) = \left(i(\delta - \Delta_1) - \frac{\kappa_1}{2}\right)a_1 + ig_m b_m^\dagger + \sqrt{\kappa_1^{\text{ext}}}\alpha_p, \tag{25}$$

$$\dot{b}_m(t) = \left(-i\delta - \frac{\Gamma_m}{2}\right)b_m + ig_m a_1^\dagger, \tag{26}$$

can be used to derive steady state values for the phonon and photon numbers

$$b_m = \frac{ig_m}{(i\delta + \frac{\Gamma_m}{2})}a_1^\dagger, \tag{27}$$

$$a_1 = \frac{-\sqrt{\kappa_1^{\text{ext}}}\alpha_p}{i(\delta - \Delta_1) - \frac{\kappa_1}{2} + \frac{g_m^2}{-i\delta + \Gamma_m/2}}. \tag{28}$$

As before, we can use the input-output formalism to obtain the transmitted probe light given by

$$a_{1,\text{out}} = -\sqrt{\kappa_1^{\text{ext}}} a_1 = \frac{\kappa_1^{\text{ext}}\alpha_p}{i(\delta - \Delta_1) - \frac{\kappa_1}{2} + \frac{g_m^2}{-i\delta + \Gamma_m/2}}. \tag{29}$$

In experiments, we use heterodyne detection of the transmitted probe light to measure the intra-cavity photon number. The power spectrum of this heterodyne signal at the detector is given by

$$P_d(\delta) \propto \langle a_{1,\text{out}}^\dagger a_{1,\text{out}} \rangle = \left| \frac{\kappa_1^{\text{ext}}}{i(\delta - \Delta_1) - \frac{\kappa_1}{2} + \frac{g_m^2}{-i\delta + \Gamma_m/2}} \right|^2. \tag{30}$$

If we assume that the frequency difference between the optical modes is exactly equal to the Brillouin frequency ($\omega_2 - \omega_1 = \Omega_m$) and the control laser is directly on resonance with the optical cavity mode $a_2$ ($\Delta_2 = 0 = \Delta_1 = 0$), we can use eqn. (45) to derive a simple expression for the intra-cavity field

$$a_1 = \frac{-\sqrt{\kappa_1^{\text{ext}}}\alpha_p}{i\delta - \frac{\kappa_1}{2} + \frac{g_m^2}{-i\delta + \Gamma_m/2}}. \tag{31}$$

Since $\Gamma_m \ll \kappa_1$ in our experiments, we get the following spectrum for the intra-cavity Stokes ($a_1$) near the phonon resonance (i.e., $\Omega \approx \Omega_m$)

$$a_1(\delta) \simeq \frac{2(i\delta - \Gamma_m/2)\sqrt{k_1^{\text{ext}}}/\kappa_1\alpha_p}{i\delta - \frac{\Gamma_m}{2}(1 - C)},$$

$$a_1(\Omega) = \frac{2(i(\Omega_m - \Omega) - \Gamma_m/2)\sqrt{k_1^{\text{ext}}}/\kappa_1\alpha_p}{i(\Omega_m - \Omega) - \frac{\Gamma_{\text{eff}}}{2}}, \tag{32}$$

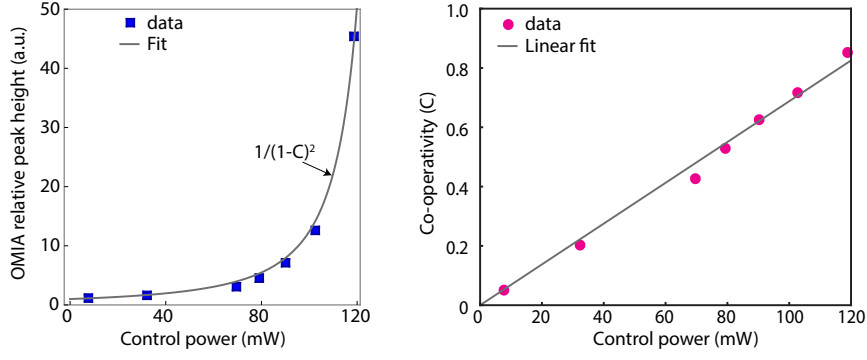

FIG. 7. **OMIA measurement a,** The relative peak height of the OMIA signal increases non-linearly with the input control laser power. This increase matches well with the theoertically predicted $1/(1-C)^2$ dependency **b,** We use the measured relative peak height of the OMIA signal to calculate multi-photon co-operativity $C$ as a function of input control laser power. We use linear fit along with the values of optical and acoustic dissipation rate in Table II to determine the single photon coupling rate $g_o^m/2\pi \simeq 18$ Hz.

where $C = 4N_2|g_0^m|^2/(\kappa_1 \Gamma_m)$ is the multi-photon co-operativity, and $\Gamma_{\text{eff}} = \Gamma_m(1 - C)$ is effective phonon linewidth.

Without optomechanical coupling (i.e. $g_m = 0$), we see that the peak intra-cavity field is given by

$$a_1'(\Omega_m) = 2\frac{\sqrt{\kappa_1^{\text{ext}}}}{\kappa_1}\alpha_p. \tag{33}$$

However, with optomechanical coupling the intra-cavity field on resonance is given by

$$a_1(\Omega_m) = 2\frac{\sqrt{\kappa_1^{\text{ext}}}}{\kappa_1(1-C)}\alpha_p \tag{34}$$

Therefore, the relative peak height of the OMIA spectrum

$$\left|\frac{a_1(\Omega_m)}{a_1'(\Omega_m)}\right|^2 = \frac{1}{(1-C)^2}, \tag{35}$$

can be used to directly measure the value of multi-photon co-operativity ($C$) in our optomechanical system.

## IV. OPTOMECHANCIALLY INDUCED TRANSPARENCY (OMIT)

In this section, we follow the approach outline in Ref. [7] to derive the intra-cavity optical spectrum during OMIT measurements. The Hamiltonain of our optomechanical system that accounts for the external drive fields (i.e., the control laser and the probe laser) is given by

$$\begin{aligned}H = &\ \hbar\omega_1 a_1^\dagger \hat{a}_1 + \hbar\omega_2 a_2^\dagger a_2 + \hbar\Omega_m b_m^\dagger b_m - \hbar g_0^m(a_2^\dagger a_1 b_m + b_m^\dagger a_1^\dagger a_2) \\ &+ i\hbar\sqrt{\kappa_1^{\text{ext}}}\alpha_l(a_1^\dagger e^{-i\omega_l t} - a_1 e^{i\omega_l t}) + i\hbar\sqrt{\kappa_2^{\text{ext}}}\alpha_p(a_2^\dagger e^{-i\omega_p t} - a_2 e^{i\omega_p t}).\end{aligned} \tag{36}$$

Note that we normalize the incident optical power launched into the cavity such that $P_l = \hbar\omega_l \langle \alpha_l^\dagger \alpha_l \rangle$. We assume a strong control laser and a weak probe; within the undepleted pump approximation, the dynamics of the mode at frequency $\omega_1$ is not influenced by the optomechanical coupling and it can be described by the following equation of motion derived from the Hamiltonian in eqn. (36)

$$\dot{a}_1(t) = \left(-i\omega_1 - \frac{\kappa_1}{2}\right)a_1 + \sqrt{\kappa_1^{\text{ext}}}\alpha_l e^{-i\omega_l t}, \tag{37}$$

where $\kappa_1 = 2\kappa_1^{\text{ext}} + \kappa_1^0$, $\kappa_1^{\text{ext}}$ is the loss rate at each cavity mirror and $k_1^0$ is the loss rate inside the cavity. From eqn. (37), we obtain the following steady state solution for mode $a_1$

$$\langle a_1 \rangle = \sqrt{N_1}e^{-i\omega_l t} = \left|\frac{\sqrt{\kappa_1^{\text{ext}}}\alpha_l}{i\Delta_1 + \kappa_1/2}\right|e^{-i\omega_l t}, \tag{38}$$

where $N_1$ is the control laser driven intra-cavity photon number for mode $a_1$, and $\Delta_1 = \omega_1 - \omega_l$. (See Fig. 6). Once we know the intra-cavity photon number for mode $a_1$, we can use input-output formalism for a symmetric Fabry-Perot cavity, assuming laser fields incident only on one of the mirrors, to obtain the transmitted light field

$$a_{1,\text{out}} = -\sqrt{\kappa_1^{\text{ext}}} a_1.$$ (39)

In the undepleted pump regime, we substitute $\langle a_1 \rangle$ in eqn. (36) to derive the following linearized Hamiltonian for the interaction between the optical mode $a_2$ and the phonon mode $b_m$

$$H = \hbar\omega_2 a_2^\dagger a_2 + \hbar\Omega_m b_m^\dagger b_m - \hbar g_0^m \sqrt{N_1}(a_2^\dagger b_m e^{-i\omega_l t} + b_m^\dagger a_2 e^{i\omega_l t}) + i\hbar\sqrt{\kappa_2^{\text{ext}}} \alpha_p (a_2^\dagger e^{-i\omega_p t} - a_2 e^{i\omega_p t}).$$ (40)

From this Hamiltonian, by rotating in the frame of $H_0 = \hbar(\omega_p - \omega_l) b_m^\dagger b_m + \hbar\omega_p a_2^\dagger a_2$, we obtain an effective Hamiltonian

$$H_{\text{eff}} = -\hbar\delta b_m^\dagger b_m - \hbar(\delta + \Delta_2) a_2^\dagger a_2 - \hbar g_m (a_2^\dagger b_m + b_m^\dagger a_2) + i\hbar\sqrt{\kappa_2^{\text{ext}}} \alpha_p (a_2^\dagger - a_2),$$ (41)

where $\delta = \Omega - \Omega_m = \omega_p - \omega_l - \Omega_a$, and $\Delta_2 = \Omega_a + \omega_l - \omega_2$, and the cavity field-enhanced coupling rate $g_m = g_0^m \sqrt{N_1}$. The Heisenberg equations of motions derived from this Hamiltonian, given by

$$\dot{a}_2(t) = \left(i(\delta + \Delta_2) - \frac{\kappa_2}{2}\right) a_2 + i g_m b_m + \sqrt{\kappa_2^{\text{ext}}} \alpha_p,$$ (42)

$$\dot{b}_m(t) = \left(i\delta - \frac{\Gamma_m}{2}\right) b + i g_m a_2,$$ (43)

can be used to derive steady state values for the phonon and photon numbers

$$b_m = \frac{-i g_m}{(i\delta - \frac{\Gamma_m}{2})} a_2,$$ (44)

$$a_2 = \frac{-\sqrt{\kappa_2^{\text{ext}}} \alpha_p}{i(\delta + \Delta_2) - \frac{\kappa_2}{2} + \frac{g_m^2}{i\delta - \Gamma_m/2}}.$$ (45)

As before, we can use the input-output formalism to obtain the transmitted probe light given by

$$a_{2,\text{out}} = -\sqrt{\kappa_2^{\text{ext}}} a_2 = \frac{\kappa_2^{\text{ext}} \alpha_p}{i(\delta + \Delta_2) - \frac{\kappa_2}{2} + \frac{g_m^2}{i\delta - \Gamma_m/2}}$$ (46)

In experiments, we use heterodyne detection of the transmitted probe light to measure the intra-cavity photon number. The power spectrum of this heterodyne signal at the detector is given by

$$P_d(\delta) \propto \langle a_{2,\text{out}}^\dagger a_{2,\text{out}} \rangle = \left| \frac{\kappa_2^{\text{ext}}}{i(\delta + \Delta_2) - \frac{\kappa_2}{2} + \frac{g_m^2}{i\delta - \Gamma_m/2}} \right|^2.$$ (47)

If we assume that the frequency difference between the optical modes is exactly equal to the Brillouin frequency ($\omega_2 - \omega_1 = \Omega_m$) and the control laser is directly on resonance with the optical cavity mode $a_1$ ($\Delta_1 = 0$), we can use eqn. (45) to derive a simple expression for the intra-cavity field

$$a_2 = \frac{-\sqrt{\kappa_2^{\text{ext}}} \alpha_p}{i\delta - \frac{\kappa_2}{2} + \frac{g_m^2}{i\delta - \Gamma_m/2}}.$$ (48)

Since $\Gamma_m \ll \kappa_2$ in our experiments, we get the following spectrum for the intra-cavity anti-Stokes (a2) near the phonon resonance (i.e., $\Omega \approx \Omega_m$)

$$a_2(\delta) \simeq \frac{2(i\delta - \Gamma_m/2)\sqrt{k_2^{\text{ext}}}/\kappa_2 \alpha_p}{i\delta - \frac{\Gamma_m}{2}(1 + C)},$$

$$a_2(\Omega) = \frac{2(i(\Omega - \Omega_m) - \Gamma_m/2)\sqrt{k_2^{\text{ext}}}/\kappa_2 \alpha_p}{i(\Omega - \Omega_m) - \frac{\Gamma_{\text{eff}}}{2}},$$ (49)

where $C = 4|g_0^m|^2 N_1/(\kappa_2 \Gamma_m)$ is the multi-photon co-operativity, and $\Gamma_{\text{eff}} = \Gamma_m(1 + C)$ is effective phonon linewidth.

Without optomechanical coupling (i.e. $g_m = 0$), we see that the peak intra-cavity field is given by

$$a_2'(\Omega_m) = 2\frac{\sqrt{\kappa_2^{\text{ext}}}}{\kappa_2}\alpha_p. \tag{50}$$

However, with optomechanical coupling the intra-cavity field on resonance is given by

$$a_2(\Omega_m) = 2\frac{\sqrt{\kappa_2^{\text{ext}}}}{\kappa_2(1 + C)}\alpha_p \tag{51}$$

Therefore, the relative dip in the OMIT spectrum which is given by

$$\left|\frac{a_2(\Omega_m)}{a_2'(\Omega_m)}\right|^2 = \frac{1}{(1 + C)^2}, \tag{52}$$

can be used to directly measure the value of multi-photon co-operativity ($C$) in our optomechanical system.

## V. ACOUSTIC DIFFRACTION LOSS

In this section, we first estimate the diffraction losses for our flat-flat resonator assuming the ideal case of no tilt misalignment between the optical cavity axis as the crystal axis. We use the approach outlined in Ref. [8], to determine diffraction losses in our system. The acoustic modes of our cylindrical resonator (See Fig. 8 a) are given by:

$$u_{m,k}(\mathbf{x}) = \beta_{m,k} \cos\left(\frac{m\pi z}{L_{\text{ac}}}\right) J_0\left(\frac{2j_{0,k}r}{d_c}\right), \tag{53}$$

where $J_0$ is the zeroth order Bessel function of the first kind, $j_{0,k}$ is the $k^{\text{th}}$ root of $J_0$, $L_{ac}$ is the thickness of the crystalline resonator and $d_c$ is the diameter of the quartz crystal, and $\beta_{m,k}$ is the normalization constant so that the total energy in each mode is equal to $\hbar\omega_{m,k}$. The frequency of each mode depends on the longitudinal mode number $m$ and the transverse mode number $k$ in the following way

$$\omega_{m,k} = \sqrt{\left(\frac{m\pi}{L_{\text{ac}}}\right)^2 v_l^2 + \left(\frac{2j_{0,k}}{d_c}\right)^2 v_t^2}, \tag{54}$$

where $v_l$ is the longitudinal sound velocity and $v_t$ is the effective transverse sound velocity. The driving of the acoustic fields in our system occurs due to the electrostrictive force, which is proportional to the square of the electric field. The optical fields for the modes of our interest are Gaussian fields with transverse area that is essentially unchanging inside the crystal. Therefore, force profile due to optical driving is also Gaussian. To calculate diffraction of the phonons that start at some initial profile $u_0(\mathbf{x}, t = 0)$ but evolve to $u(\mathbf{x}, t)$ after some time $t$, we use the eigenmode expansion as

$$u(\mathbf{x}, t) = \sum_{m',k'} c_{m',k'} e^{-i\omega_{m',k'}t} u_{m',k'}(\mathbf{x}), \tag{55}$$

where the constant $c_{m',k'}$s are determined from the initial displacement profile due to the electrostrictive driving given by

$$u_0(\mathbf{x}, t = 0) = U_0 \cos\left(\frac{m\pi z}{L_{ac}}\right) e^{-2r^2/w_o^2}, \tag{56}$$

where $\text{w}_o$ is the waist of the optical Gaussian beam that drives the phonon modes. To calculate diffraction losses we calculate the overlap of the phonon field at time $t$ with the original state $u_0(\mathbf{x}, t = 0)$. In particular we look at

$$|\eta(t)|^2 = |\langle u(\mathbf{x}, t)|u(\mathbf{x}, t = 0)\rangle|^2 \tag{57}$$

$$= \left|\sum_{k'} |c_{m,k'}|^2 e^{i\omega_{m,k'}t}\right|^2. \tag{58}$$

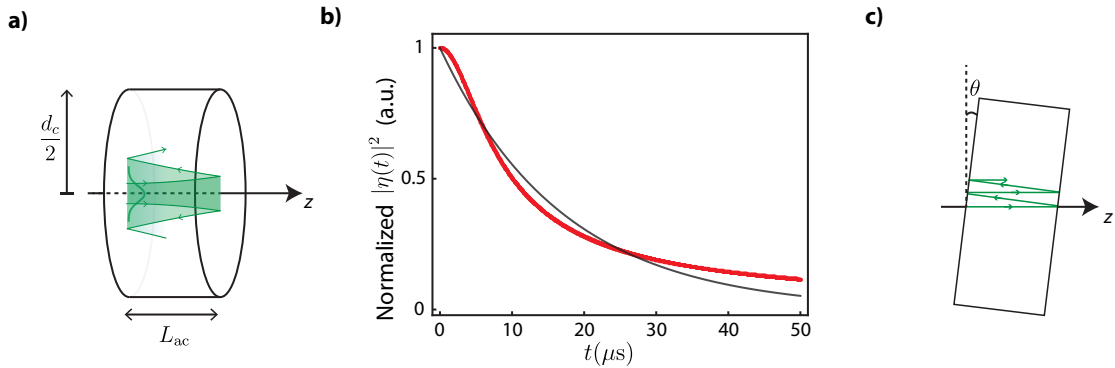

FIG. 8. **Acoustic diffraction loss. a,** Our acoustic resonator has a cylindrical geometry, with $L_{ac} = 5.185$ mm and $d_c = 12.7$ mm. Gaussian acoustic beam diffracts upon propagation inside the crystalline medium resulting in loss of acoustic energy. **b,** We obtain a diffraction loss limited lifetime of $\sim 17 \mu$s, which corresponds to a phonon linewidth of $\sim 9$ kHz. For this calculation of acoustic diffraction loss we used the following parameters: $w_o = 43$ μm, $v_l = 6319$ m/s, $v_t = 5112$ m/s, and $\Omega_m/2\pi = 12.645$ GHz. **c,** Even a small tilt angle ($\theta$) of the crystal axis with respect to the acoustic axis can result in a large spatial walk-off of the phonon beam resulting in diffractive losses.

We obtain the acoustic energy lost from diffraction by plotting the normalized value of $|\eta(t)|^2$ (i.e. $|\eta(t)|^2/|\eta(t = 0)|^2$) as a function of time (See Fig. 8 b). Fitting an exponential to this plot, we get an order of magnitude estimate for the decay time $\sim 17 \mu s$, which corresponds to phonon linewidth $\Gamma_m/2\pi = 9$ kHz.

In experiments, however, we measure a linewidth ($\Gamma_m/2\pi = 86$ kHz) that is much larger than the diffraction loss limit of 9 kHz. Note that for the diffraction loss calculation, we assumed that there is no tilt misalignment between the optical cavity axis and the crystal axis. Phonons encounter additional diffractive losses in the planar acoustic geometry if the crystal axis is tilted with respect to the optical cavity axis (See Fig. 8 c). This is analogous to the well known losses introduced by tilting a Fabry-Perot etalon inside an optical cavity [9].

The measured finesse of our acoustic cavity was about 7. We calculate that even a small tilt angle of 0.07 degree is enough to result in lateral offset for the Gaussian acoustic beam that is twice the initial acoustic waist after 7 roundtrips. Since the crystal is sitting on the Invar holder, whose surfaces are not perfectly flat, it is entirely possible in our experiment to have this level of misalignment. By designing a stable plano-convex acoustic cavity it is possible to mitigate the effects of diffraction and make the acoustic loss less sensitive to tilt misalignment.

Acoustic $Q$-factor of $4.2 \times 10^7$ ($\Gamma_m \approx 2\pi \times 300$ Hz) has already been demonstrated for 12.7 GHz phonons confined within such plano-convex quartz resonators at cryogenic temperatures ($\sim 10$ K) [6]. It is intriguing to consider using such highly coherent phonons to access, control, and study defect centers (or Two Level Systems (TLS)) in a variety of materials [10, 11]. Using such high-Q resonators fabricated on-chip [12], it is possible to create phonon modes with mode volumes of $\sim 1.6 \times 10^{-13}$ having zero-point strain fields of $1.5 \times 10^{-11}$. Assuming a deformation potential of 1 eV per unit strain for a single TLS, we estimate zero-point strain induced frequency shift of TLS energy level of about 4 kHz. Note than this zero-point coupling rate for phonon-TLS interaction is more than 10 times larger than acoustic cavity linewidth. Therefore, utilizing defect centers that have very long lifetimes at cryogenic temperatures [13], it may be possible to enter the regime of strong coupling regime for phonon-TLS interaction. Such regime would allow phononic access to defect centers for scientific studies of phononic decoherence mechanisms as well as for quantum acoustic technologies.

## VI. THERMAL FLUCTUATIONS AND PHONON LASING

In this section, we look at the power spectrum of the spontaneously scattered Stokes light due to thermal fluctuations of our mechanical mode. We consider the case where the control laser in directly on resonance with the high-frequency optical mode ($\omega_l = \omega_2$) and the frequency spacing between the two adjacent optical modes is equal to the Brillouin frequency ($\omega_2 - \omega_1 = \Omega_m$).

For measurements of thermal fluctuations we turn off the probe laser ($\alpha_p = 0$) while the control laser is directly on resonance with the higher frequency optical mode at $\omega_2$. Starting with the Hamiltonian from eqn. (19) and setting

$\alpha_p = 0$, we obtain

$$H = \hbar\omega_1 a_1^\dagger \hat{a}_1 + \hbar\omega_2 a_2^\dagger a_2 + \hbar\Omega_m b_m^\dagger b_m - \hbar g_0^m (a_2^\dagger a_1 b_m + b_m^\dagger a_1^\dagger a_2)$$
$$+ i\hbar\sqrt{\kappa_2^{\text{ext}}}\alpha_l (a_1^\dagger e^{-i\omega_2 t} - a_2 e^{i\omega_2 t}). \tag{59}$$

As before, we assume a strong control laser and weak optomechanical coupling to derive the following steady state amplitude for mode $a_2$

$$\langle a_2 \rangle = \sqrt{N_2} e^{-i\omega_l t} = 2\frac{\sqrt{\kappa_2^{\text{ext}}}}{\kappa_2}\alpha_l e^{-i\omega_2 t}, \tag{60}$$

where $N_2$ is the intra-cavity photon number in mode $a_2$. We substitute $\langle a_2 \rangle$ into the Hamiltonian in eqn. (59) and by rotating in the frame of the control laser (i.e. $H_0 = \hbar\omega_2 a_2^\dagger a_2$), we obtain the following linearized Hamiltonian

$$H_{\text{eff}} = -\hbar\Omega_m a_1^\dagger a_1 + \hbar\Omega_m b_m^\dagger b_m - \hbar g_m (a_1 b_m + b_m^\dagger a_1^\dagger), \tag{61}$$

where $g_m = \sqrt{N_2} g_0^m$ is the cavity-enhanced optomechanical coupling rate. The Heisenberg-Langevin equations of motion derived from this effective Hamiltonian are

$$\dot{a}_1 = \left(i\Omega_m - \frac{\kappa_1}{2}\right) a_1 + i g_m b_m^\dagger \tag{62}$$

$$\dot{b}_m = \left(-i\Omega_m - \frac{\Gamma_m}{2}\right) b_m + i g_m a_1^\dagger + \sqrt{\Gamma_m}\tilde{b}_{\text{in}} \tag{63}$$

where $\tilde{b}_{\text{in}}(t)$ is the input thermal fluctuation that drives the phonon mode. We assume a Markovian noise process [14, 15] such that

$$\langle \tilde{b}_{\text{in}}^\dagger(t)\tilde{b}_{\text{in}}(t') \rangle = n_{\text{th}}\delta(t - t') \tag{64}$$

$$\langle \tilde{b}_{\text{in}}(t)\tilde{b}_{\text{in}}^\dagger(t') \rangle = (n_{\text{th}} + 1)\delta(t - t'). \tag{65}$$

$\langle \ldots \rangle$ represents an ensemble average, and $n_{\text{th}} = (e^{\hbar\Omega_m/k_B T} - 1)^{-1}$ denotes the average number of thermal phonons of angular frequency $\Omega_m$ at temperature $T$. For instance, at a temperatures of 10 Kelvin, $n_{\text{th}} \simeq 16$ for 12.645 GHz phonon modes. In contrast, the optical fields have essentially zero thermal occupation even at room temperatures since they are at much higher frequencies ($\sim 200$ THz). Therefore, we ignore thermal fluctuations of the optical field in our calculations. Furthermore, we assume that the externally driven laser source is a pure coherent tone (no added noise from the laser itself), which is a good approximation for our experiments.

We now use Fourier-transform (defined as $f(\omega) = \int_{-\infty}^{+\infty} dt\, f(t) e^{i\omega t}$) to solve eqns. (62-63) in frequency space, which are given by

$$a_1(\omega) = \frac{-i g_m}{i(\omega + \Omega_m) - \kappa_1/2} b_m^\dagger(\omega) \tag{66}$$

$$b_m(\omega) = \frac{i\sqrt{\Gamma_m}}{(\omega - \Omega_m + i\Gamma_m/2 + \Sigma(\omega))}\tilde{b}_{\text{in}}(\omega), \tag{67}$$

where we define

$$\Sigma(\omega) = \delta\Omega_m(\omega) - i\frac{\Gamma_{\text{opt}}(\omega)}{2} = \frac{i g_m^2}{i(\omega - \Omega_m) - \kappa_1/2}. \tag{68}$$

Here, $\delta\Omega_m(\omega) = \text{Re}\{\Sigma(\omega)\}$ gives us the frequency shift of the phonon mode and $\Gamma_{\text{opt}}(\omega) = -2\text{Im}\{\Sigma(\omega)\}$ gives us the optomechanical amplification rate. While $\Sigma(\omega)$ is frequency dependent, for weak coupling ( $g_m \ll \kappa_1$), we can just evaluate $\delta\Omega_m(\omega)$ and $\Gamma_{\text{opt}}(\omega)$ at $\omega = \Omega_m$ [14]. This calculation gives $\delta\Omega_m = 0$ and $\Gamma_{\text{opt}} = 4g_m^2/\kappa_1$. So, we can re-write eqn. (67) as

$$b_m(\omega) = \frac{i\sqrt{\Gamma_m}}{(\omega - \Omega_m + i\Gamma_m'/2)}\tilde{b}_{\text{in}}(\omega), \tag{69}$$

where $\Gamma_m' = \Gamma_m - \Gamma_{\text{opt}} = \Gamma_m - 4g_m^2/\kappa_1$

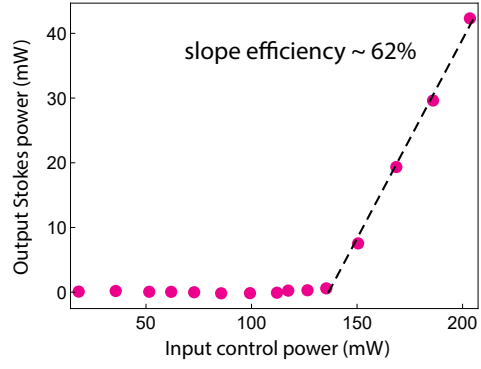

FIG. 9. **Slope efficiency of the phonon laser**. The backscattered optical power is used to determine total output Stokes power as a function of the input control laser power. This measurement reveals a slope efficiency of $\approx 62\%$, which is consistent with the theoretically estimated value of 57%.

We can use $b(\omega)$ in eqn. (69) along with the noise correlations in eqns.(64)-(65) to calculate the power spectrum of laser driven mechanical mode, $S_{bb}(\omega)$, which is defined as

$$S_{bb}(\omega) = \int_{-\infty}^{+\infty} \langle b^\dagger(t)b(0) \rangle e^{i\omega t} dt \tag{70}$$

$$= \frac{\Gamma_m n_{\mathrm{th}}}{(\omega + \Omega_m)^2 + (\Gamma'_m/2)^2}. \tag{71}$$

From this derivation we see that the power spectrum has a linewidth determined by the effective damping rate $\Gamma'_m = \Gamma_m(1 - \Gamma_{\mathrm{opt}}/\Gamma_m)$. Therefore, when the optomechanical amplification rate equals the cold cavity linewidth ($\Gamma_{\mathrm{opt}} = \Gamma_m$ or $C = \Gamma_{\mathrm{opt}}/\Gamma_m = 1$), a threshold condition is achieved, which leads to zero effective damping and regenerative self-oscillation of the phonon mode (also called phonon lasing).

We can also calculate the intra-cavity power spectrum of the spontaneously scattered light due to thermal fluctuations of the mechanical mode using eqns. (66-67). The power spectrum, $S_{a_1 a_1}(\omega)$, is given by

$$S_{a_1 a_1}(\omega) = \int_{-\infty}^{+\infty} \langle a_1^\dagger(t)a_1(0) \rangle e^{i\omega t} dt \tag{72}$$

$$= \frac{g_m^2 \Gamma_m (n_{\mathrm{th}} + 1)}{((\omega - \Omega_m)^2 + (\kappa_1/2)^2)((\omega - \Omega_m)^2 + (\Gamma'_m/2)^2)}. \tag{73}$$

From this expression we see that, using proper calibration of the heterodyne power spectrum of the spontaneously scattered light and values of $g_m$, $\Gamma_m$, $\kappa_1$ determined independently from the coherent (OMIA) measurements, it is possible to estimate the thermal phonon number in our system.

### A. Slope efficiency

Once self-oscillating (i.e. above threshold) the total Stokes power as a function of input control power is given by [16]

$$P_s \approx \frac{\kappa_p \omega_s v_p}{2\kappa_s \omega_p v_s}(P_{\mathrm{in}} - P_{\mathrm{th}}), \tag{74}$$

where $\kappa_p(\kappa_s)$ is the optical cavity linewidth, $v_p(v_s)$ is the group velocity of optical mode at frequency $\omega_p(\omega_s)$ respectively, and $P_{\mathrm{th}}$ is the threshold input power for self-oscillation. Using the parameters from table II we calculate the slope efficiency $P_s/(P_{\mathrm{in}} - P_{\mathrm{th}}) \approx 57\%$. This agrees well with the experimentally measured value of 62 % (See Fig. 9).

### B. Phonon laser linewidth

Next we estimate the quantum-back-action limited linewidth and phase noise of our phonon laser. At $C = 1.4$, from the measurements of backreflected optical power, we estimate total scattered output Stokes power of $P_s = 42$

mW (See Fig. 9). Knowing the scattering rate of Stokes light per phonon, $\gamma_{\mathrm{OM}} = 4G^2/\kappa$, we can calculate the steady state coherent phonon number as

$$n_c = \frac{P_s}{\hbar\omega\gamma_{\mathrm{OM}}} = 4.2 \times 10^{11}. \tag{75}$$

This large coherent phonon number should produce dramatic Schawlow-Townes narrowed linewidth [17] given by

$$\Delta\Omega = \frac{\Gamma_m}{2n_c}(n_{\mathrm{th}} + 1) = 2\pi \times 1.7 \ \mu\mathrm{Hz}, \tag{76}$$

where $n_{\mathrm{th}} = 16$ at 10 Kelvin temperatures. Such ultra-narrow linewidth would result in a phase noise of -145 dBc/Hz at 10 KHz offset for a 12.6 GHz opto-mechanical oscillator [18]. Dramatic improvement in the quantum-back-action limited phase noise performance is possible within this quartz optomechanical system by using a plano-convex acoustic resonator having an intrinsic phonon disspation rate $\Gamma_m = 2\pi \times 300$ Hz [19]. For similar control input laser powers of 204 mW, we obtain steady state phonon population of $2 \times 10^{12}$. This would result in a Schawlow-Townes narrowed linewidth of $\Delta\Omega = 2\pi \times 0.5$ nHz at 4 Kelvin temperature. Such highly coherent oscillator would have a phase noise of -181 dBc/Hz at 10 kHz offset for a 12.7 GHz opto-mechanical oscillator.

## VII. RELATIVE SCATTERING RATE

In this section, we calculate the scattering rate difference between the Stokes and anti-Stokes processes for our multimode optomechanical system. In a single mode optomechanical system, a control laser is typically detuned from an optical resonance by $\Omega_m$ to achieve a large difference in the Stokes and the anti-Stokes scattering rate (provided that $\Omega_m \gg \kappa$). However, the coherent driving field is directly on resonance with an optical mode in our optomechanical system. Nevertheless, we can engineer large asymmetry in the scattering rates for the Stokes and the anti-Stokes processes by engineering the optical density of states (See Fig. 10a).

To explore this we consider a Hamiltonian that describes three optical modes interacting with a single phonon mode

$$H = \sum_{j=0,1,2} \hbar\omega_j a_j^\dagger a_j + \hbar\Omega_m b_m^\dagger b_m - \hbar g_{0,-}^m(a_1^\dagger a_0 b_m + b_m^\dagger a_0^\dagger a_1) - \hbar g_{0,+}^m(a_2^\dagger a_1 b_m + b_m^\dagger a_1^\dagger a_2) + H_{\mathrm{drive}}. \tag{77}$$

Here, $a_j$ ($b_m$) is the annihilation operator for the optical (phonon) mode at frequency $\omega_j$ ($\Omega_m$), $g_j^m$ is the single photon coupling rate for the Stokes ( anti-Stokes) scattering processes, and $H_{\mathrm{drive}}$ is the Hamiltonian for the external drive field. Since all the fundamental longitudinal optical modes considered here have essentially the same mode profiles $g_{0,+} = g_{0,-} = g_0^m$ is an excellent approximation.

For our calculation, we assume that the coherent laser field is driving the optical mode at frequency $\omega_1$ directly on resonance ($\omega_l = \omega_1$). We also assume that only two optical modes ($a_1$ and $a_2$) are frequency separated by the phonon frequency ($\omega_2 - \omega_1 = \Omega_m$). Then, $\omega_1 - \omega_0 = \Omega_m + \chi$, where $\chi$ characterizes the asymmetry in the optical FSR. We also assume that the decay rate of the optical modes is due to the external coupling loss due to finite reflectivity of the two mirrors ($\kappa_j = 2\kappa_{\mathrm{ext}}$). For weak optomechanical coupling, we assume undepleted pump and a large coherent field amplitude $\langle a_1 \rangle = \sqrt{N_1}e^{-i\omega_1 t}$, where $\sqrt{N_1} = 2(\sqrt{\kappa_{\mathrm{ext}}}/\kappa)\alpha_l$ and $P_{\mathrm{in}} = \hbar\omega_1\langle\alpha_l^\dagger\alpha_l\rangle$.

Working in the rotating frame of $H_0 = \hbar\omega_1 a_1^\dagger a_1$, we obtain the following linearized Hamiltonian from eqn. (77)

$$H_{\mathrm{eff}} = -\hbar(\Omega_m + \chi)a_0^\dagger a_0 + \hbar\Omega_m a_2^\dagger a_2 - \hbar g_m(a_0 b_m + b_m^\dagger a_0^\dagger) - \hbar g_m(a_2^\dagger b_m + b_m^\dagger a_2), \tag{78}$$

where $g_m = g_0^m\sqrt{N_1}$. We use this Hamiltonian to derive the following Heisenberg-Langevin equations of motion for the dynamics of the anti-Stokes photon, Stokes photon and the phonon

$$\dot{b}_m = \left(-i\Omega_m - \frac{\Gamma_m}{2}\right)b_m(t) + ig_m a_0^\dagger + ig_m a_2 + \sqrt{\Gamma_m}\tilde{b}_{\mathrm{in}}, \tag{79}$$

$$\dot{a}_0 = \left(i(\Omega_m + \chi) - \frac{\kappa_0}{2}\right)a_0 + ig_m b_m^\dagger, \tag{80}$$

$$\dot{a}_2 = \left(-i\Omega_m - \frac{\kappa_2}{2}\right)a_2(t) + ig_m b_m. \tag{81}$$

Using these equations we obtain the following equation for the phonon mode amplitude in frequency space:

$$b_m(\omega) = \frac{i\sqrt{\Gamma_m}\tilde{b}_{\mathrm{in}}(\omega)}{(\omega - \Omega_m) + i\Gamma_m/2 + \Sigma^-(\omega) + \Sigma^+(\omega)}, \tag{82}$$

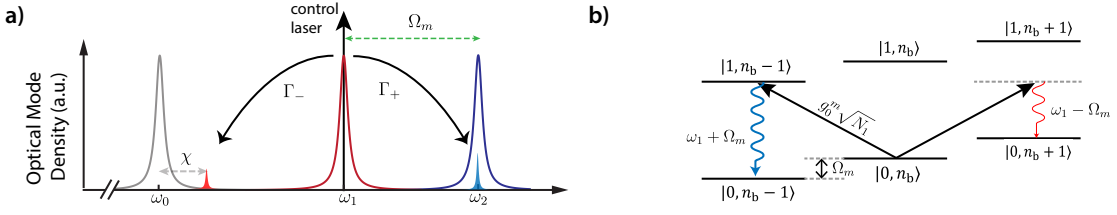

FIG. 10. **asymmetry in the Stokes and anti-Stokes scattering processes a**, A strong control laser is driven on resonance with the mode at frequency $\omega_1$. A strong suppression of the Stokes scattering (or phonon mode heating) occurs relative to the anti-Stokes scattering (or phonon mode cooling) due to unequal FSR. **b**, Energy level diagram depicting optomechanical damping and amplification of phonons.

where the modification to the mechanical susceptibility due to optomechanical coupling is given by

$$\Sigma^-(\omega) = \delta\Omega_m^-(\omega) + i\frac{\Gamma_-(\omega)}{2} = \frac{ig_m^2}{i(\omega - \Omega_m - \chi) - \kappa_0/2}, \tag{83}$$

$$\Sigma^+(\omega) = \delta\Omega_m^+(\omega) + i\frac{\Gamma_+(\omega)}{2} = \frac{-ig_m^2}{i(\omega - \Omega_m) - \kappa_0/2}. \tag{84}$$

As discussed in Section VI, the real and imaginary parts of $\Sigma^\pm(\omega)$ determine mechanical frequency shift ($\delta\Omega_m^\pm(\omega)$) and the optomechanical damping/amplification rate ($\Gamma_\pm(\omega)$). For weak coupling $g_m \ll \kappa/2$ and $\Gamma_m \ll \kappa$, we can simply evaluate $\Sigma^\pm(\omega)$ at the mechanical frequency $\omega = \Omega_m$ to get

$$\delta\Omega_m^- = -\frac{g_m^2\chi}{\chi^2 + (\kappa_0/2)^2}, \tag{85}$$

$$\Gamma_- = -\frac{g_m^2\kappa_0}{\chi^2 + (\kappa_0/2)^2}, \tag{86}$$

$$\delta\Omega_m^- = 0, \tag{87}$$

$$\Gamma_+ = \frac{4g_m^2}{\kappa_2}. \tag{88}$$

If we assume $\kappa_0 \simeq \kappa_2 = \kappa$ and $\chi \gg \kappa$, we obtain

$$\Gamma_- \simeq = -4\frac{g_m^2\kappa}{\chi^2} = -\Gamma_+ \left(\frac{\kappa}{2\chi}\right)^2. \tag{89}$$

Therefore, for the optomechanical cooling process that we considered here (See Fig. 10b), the Stokes process (phonon mode heating) is smaller than the anti-Stokes process (phonon mode cooling) by a factor of $(\kappa/2\chi)^2$.

## VIII. PHONON COUNTING SENSITIVITY

In this section, we derive phonon counting sensitivity in the weak coupling regime ($g_m \ll \kappa$) for our multi-mode optomechanical system and compare it with the phonon counting sensitivity for a single-mode optomechanical system presented in Ref. [20].

We use a strong coherent drive field that is directly resonant on the lower frequency optical mode $\omega_j$ while the probe laser ($\alpha_p = 0$) is turned off (See Fig. 11a). We assume that $\omega_{j+1} - \omega_j = \Omega_m$. We use the following linearized Hamiltonian

$$H^{\text{eff}} = \hbar\omega_{j+1}a_{j+1}^\dagger a_{j+1} + \hbar\Omega_m b_m^\dagger b_m - \hbar G(a_{j+1}^\dagger b_m e^{-i\omega_l t} + b_m^\dagger a_{j+1}e^{i\omega_l t}) \tag{90}$$

to obtain Heisenberg equations of motion and use the input-output relations, to get the following relation for the

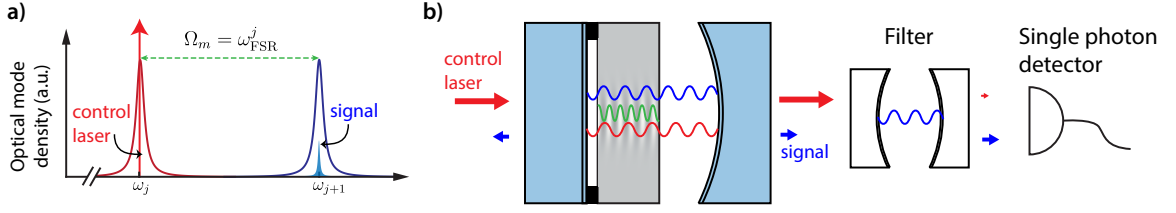

FIG. 11. **Phonon counting in a multi-resonant optomechanical system a**, A strong control laser is driven on with a low frequency optical mode at frequency $\omega_j$ for phonon couting measurements. **b**, The transmitted light from our optomechanical system consists of both the signal (light scattered due to phonons) as well as the strong control laser, which must be filtered out before photon counting measurements.

high-frequency optical field exiting our optomechanical system

$$a_{\text{out}}(t) = -\sqrt{\kappa_e}a_{j+1}(t) \tag{91}$$

$$\simeq \frac{2i\sqrt{\kappa_e}Gb_m(t)}{\kappa} \tag{92}$$

$$= i\sqrt{\frac{\kappa_e}{\kappa}}\sqrt{\gamma_{\text{OM}}}b_m(t), \tag{93}$$

where $G = \sqrt{n_c}g_0^m$ is the cavity enhanced coupling rate, $\kappa_e$ is the external loss rate at each cavity mirror, $\kappa = 2\kappa_e$ is the total decay rate for each optical mode, and the optomechanical damping rate $\gamma_{\text{OM}} = 4G^2/\kappa$. Since, $\langle a_{\text{out}}^\dagger a_{\text{out}}\rangle = (\kappa_e/\kappa)\gamma_{\text{OM}}\langle b_m^\dagger b_m\rangle$, the rate of emission photons at frequency $\omega_{j+1}$ per phonon is $(\kappa_e/\kappa)\gamma_{\text{OM}}$. These photons are detected with an overall system detection efficiency of $\eta$ (this includes losses before the photon gets to the single photon detector as well as the less than unity detection efficiency of the single photon detector). The count rate per phonon on the detector due to our signal is $\Gamma_{\text{sig}} = \eta(\kappa_e/\kappa)\gamma_{\text{OM}}$.

In contrast the single-mode optomechanical system, the light transmitted through our optomechanical systems also has a large flux of control photons because the control laser is directly on resonance with the optical mode at $\omega_j$. Assuming critical coupling and no-internal losses, photon flux for the transmitted control laser in terms of the intra-cavity photon number is given by

$$\dot{N}_c = \frac{P_{\text{in}}}{\hbar\omega} = \frac{n_c\kappa^2}{4\kappa_e}. \tag{94}$$

The transmitted light is passed through an additional optical filter to remove unwanted control laser (See Fig. 11b) before photon detection. Let's assume an power attenuation of $A$. Then the count rate on the detector due to the wanted control laser is $\Gamma_{\text{control}} = \eta A\dot{N}_c$.

So, the total count rate on the detector is given by

$$\Gamma_{\text{tot}} = \Gamma_{\text{sig}}\langle n\rangle + \Gamma_{\text{control}} + \Gamma_{\text{dark}}, \tag{95}$$

where $\langle n\rangle$ is the average phonon number and $\Gamma_{\text{dark}}$ is the dark count rate of the signal photon detector.

To characterize the sensitivity of phonon counting we calculate the noise-equivalent phonon number for our multi-mode optomechanical system

$$n_{\text{NEP}}^{\text{MM}} = n_{\text{control}}^{\text{MM}} + n_{\text{dark}}^{\text{MM}} \tag{96}$$

$$= \frac{\Gamma_{\text{control}} + \Gamma_{\text{dark}}}{\Gamma_{\text{sig}}} \tag{97}$$

$$= A\left(\frac{\kappa^2}{4\kappa_e g_0}\right)^2 + \frac{\Gamma_{\text{dark}}\kappa^2}{4\eta\kappa_e n_c g_0^2} \tag{98}$$

For single phonon level sensitivity, we want $n_{\text{NEP}} < 1$. In comparison, the noise-equivalent phonon number for a single-mode optomechanical system is given by [20]

$$n_{\text{NEP}}^{\text{SM}} = n_{\text{control}}^{\text{SM}} + n_{\text{dark}}^{\text{SM}} \tag{99}$$

$$= A\left(\frac{\kappa\Omega_m}{2\kappa_e g_0}\right)^2 + \frac{\Gamma_{\text{dark}}\kappa^2}{4\eta\kappa_e n_c g_0^2} \tag{100}$$

For same values of $n_c$, the expression for $n_{\text{dark}}$ is the same for both multi-mode and single-mode optomechanical system. At experimentally achievable $\gamma_{\text{OM}} > 1$ MHz and dark count rates of $\Gamma_{\text{dark}} = 10$ Hz, $n_{\text{dark}} \ll 1$.

Assuming symmetric Fabry-Pèrot type optomechanical system ($\kappa = 2\kappa_e$), we find from eqn. (96) and (99) that

$$n_{\text{control}}^{\text{SM}} = A \left( \frac{\Omega_m}{g_0} \right)^2 \tag{101}$$

$$n_{\text{control}}^{\text{MM}} = A \left( \frac{\kappa}{2g_0} \right)^2 \tag{102}$$

From this we see that $n_{\text{control}}^{\text{MM}}$ is suppressed by the square of the optical finesse. However, $n_{\text{control}}^{\text{SM}}$ in a single mode optomechanical system is independent of the optical finesse. Instead $n_{\text{control}}^{\text{SM}}$ increases as the square of the frequency of the phonon mode.

Note that the calculation we performed here is a conservative estimate because sensitivity can be enhanced by performing photon counting using the back-reflected light. When the control laser is critically coupled to our optomechanical cavity, back-reflected control laser power, in principle, is zero. In practice, we expect around 20 dB suppression of the unwanted counts from the control laser when using the back-reflection port.

Using a higher finesse optical cavity ($\mathcal{F} \approx 3000$) with $\kappa/2\pi \approx 4$ MHz and $g_0/2\pi = 20$ Hz, we expect $n_{\text{control}}^{\text{MM}} < 1$ using 80 dB of filtering of the control laser in our multi-mode optomechanical system. In this system we get $\gamma_{\text{OM}} = 2\pi \times 2$ MHz for $P_{\text{in}} \approx 11$ mW. Assuming $\eta = 50\%$, we get $\Gamma_{\text{sig}} = 2\pi \times 0.5$ MHz.

---


\newpage\null\thispagestyle{empty}\newpage
\end{document}